\documentclass[aps,prd,10pt,twocolumn,superscriptaddress,floatfix,nofootinbib,amsmath,amssymb,altaffilletter,preprintnumbers]{revtex4-1}
\usepackage[colorlinks=true,pagebackref=true,pdfstartview=FitV,breaklinks=true]{hyperref}

\usepackage{times}
\usepackage{graphicx}
\usepackage{dcolumn}
\usepackage{bm}
\usepackage{tensor}
\usepackage{slashed}
\usepackage{bm}
\usepackage{multirow}
\usepackage{soul}
\usepackage{amsmath}
\usepackage{fontawesome}
\usepackage{mathrsfs}
\usepackage{amssymb}
\usepackage{yfonts}
\usepackage{color}
\usepackage{xspace}
\usepackage{url}
\usepackage{verbatim}
\usepackage{mathtools}
\usepackage{upgreek}
\usepackage{amstext}
\usepackage{booktabs}
\usepackage{tabulary}
\usepackage{tabularx}
\usepackage{etoolbox}
\usepackage[utf8]{inputenc}
\usepackage[dvipsnames,table]{xcolor}
\newcolumntype{Y}{>{\centering\arraybackslash}X}
\hypersetup{urlcolor=BlueViolet,
	    citecolor=Plum,
	    linkcolor=PineGreen}
\usepackage[official]{eurosym}
\definecolor{lightgray}{rgb}{0.9,0.9,0.9}	    
\definecolor{green}{rgb}{0,0.5,0}
\definecolor{red}{rgb}{1,0,0}
\definecolor{blue}{rgb}{0,0,0.5}

\newcommand{\qhat}{\hat{\mathbf{q}}}

\newcommand{\dbd}[2]{\ifmmode \frac{\textrm{d}#1}{\textrm{d}#2}\else $\textrm{d}#1/\textrm{d}#2$\fi}
\newcommand{\pbp}[2]{\ifmmode \frac{\partial#1}{\partial#2}\else $\partial#1/\partial#2$\fi}

\newcommand{\drm}{\mathrm{d}}

\DeclareMathAlphabet{\mathpzc}{OT1}{pzc}{m}{it}
\newcommand{\eV}{\text{e\kern-0.15ex V}\xspace}
\newcommand{\keV}{\text{k\eV}\xspace}

\newcommand{\TeV}{\text{T\kern-0.1ex \eV}\xspace}

\newcommand{\cevns}{CE$\nu$NS\xspace}
\newcommand{\vees}{$\nu e$ES\xspace}
 \newcommand{\keVr}{\text{k\text{e\kern-0.15ex V}$_\mathrm{r}$}\xspace}

\newcommand{\Cygnus}{\textsc{Cygnus}\xspace}

\newcommand{\Boron}{$^8$B\xspace}
\newcommand{\be}{\begin{equation}}
\newcommand{\ee}{\end{equation}}
\newcommand{\bea}{\begin{eqnarray}}
\newcommand{\eea}{\end{eqnarray}}

\begin{document}

\title{CYG$\nu$S: Detecting solar neutrinos with directional gas time projection chambers}

\author{Chiara Lisotti}
\email{maria.lisotti@sydney.edu.au}
\affiliation{ARC Centre of Excellence for Dark Matter Particle Physics, The University of Sydney, School of Physics, NSW 2006, Australia}

\author{Ciaran A.~J.~O'Hare}
\email{ciaran.ohare@sydney.edu.au}
\affiliation{ARC Centre of Excellence for Dark Matter Particle Physics, The University of Sydney, School of Physics, NSW 2006, Australia}

\author{Elisabetta~Baracchini}
\affiliation{Department of Astroparticle Physics, Gran Sasso Science Institute, L’Aquila, I-67100, Italy}
\affiliation{Istituto Nazionale di Fisica Nucleare, Laboratori Nazionali del Gran Sasso, 67100 Assergi, Italy}

\author{Victoria~U.~Bashu}
\affiliation{ARC Centre of Excellence for Dark Matter Particle Physics, Research School of Physics, Australian National University, ACT 2601, Australia}

\author{Lindsey~J.~Bignell}
\affiliation{ARC Centre of Excellence for Dark Matter Particle Physics, Research School of Physics, Australian National University, ACT 2601, Australia}

\author{Ferdos~Dastgiri}
\affiliation{ARC Centre of Excellence for Dark Matter Particle Physics, Research School of Physics, Australian National University, ACT 2601, Australia}

\author{Majd~Ghrear}
\affiliation{Department of Physics and Astronomy, University of Hawaii,
2505 Correa Road, Honolulu, HI, 96822, USA}

\author{Gregory~J.~Lane}
\affiliation{ARC Centre of Excellence for Dark Matter Particle Physics, Research School of Physics, Australian National University, ACT 2601, Australia}

\author{Lachlan~J.~McKie}
\affiliation{ARC Centre of Excellence for Dark Matter Particle Physics, Research School of Physics, Australian National University, ACT 2601, Australia}

\author{Peter C.~McNamara}
\affiliation{ARC Centre of Excellence for Dark Matter Particle Physics$,$ \\~School of Physics$,$~ The~ University~ of~ Melbourne$,$~ Victoria~ 3010$,$~ Australia}

\author{Samuele~Torelli}
\affiliation{Department of Astroparticle Physics, Gran Sasso Science Institute, L’Aquila, I-67100, Italy}
\affiliation{Istituto Nazionale di Fisica Nucleare, Laboratori Nazionali del Gran Sasso, 67100 Assergi, Italy}

\smallskip
\begin{abstract}
\Cygnus is a proposed global network of large-scale gas time projection chambers (TPCs) with the capability of directionally detecting nuclear and electron recoils at $\gtrsim$keV energies. 
The primary focus of \Cygnus so far has been the detection of dark matter, with directional sensitivity providing a means of circumventing the so-called ``neutrino fog''.
However, the excellent background rejection and electron/nuclear recoil discrimination provided by directionality could turn the solar neutrino background into an interesting signal in its own right. For example, directionality would facilitate the simultaneous spectroscopy of multiple different flux sources.
Here, we evaluate the possibility of measuring solar neutrinos using the same network of gas TPCs built from $10$~m$^3$-scale modules operating under conditions that enable simultaneous sensitivity to both dark matter and neutrinos.
We focus in particular on electron recoils, which provide access to low-energy neutrino fluxes like $pp$, $pep$, $^7$Be, and CNO. An appreciable event rate is already detectable in experiments consisting of a single $10$~m$^3$ module, assuming standard fill gases such as CF$_4$ mixed with helium at atmospheric pressure.
With total volumes around 1000 m$^3$ or higher, the TPC network could be complementary to dedicated neutrino observatories, whilst entering the dark-matter neutrino fog via the nuclear recoil channel.
We evaluate the required directional performance and background conditions to observe, discriminate, and perform spectroscopy on neutrino events. We find that, under reasonable projections for planned technology that will enable 10--30-degree angular resolution and $\sim 10$\% fractional energy resolution, \Cygnus could be a competitive directional neutrino experiment.
\end{abstract}

\maketitle

\section{Introduction}
Over the last three decades, a major driving force in the development of ultralow-background underground particle detectors has been the search for interactions of galactic dark matter with Standard Model particles~\cite{Battaglieri:2017aum,Schumann:2019eaa}. What has been anticipated for some years~\cite{Monroe:2007xp,Vergados:2008jp,Strigari:2009bq,Gutlein:2010tq}, but is only now becoming a reality, is an era in which detectors intended to search for dark matter-induced recoils are large enough to serve a dual purpose as detectors of astrophysical neutrinos ~\cite{Harnik:2012ni,Pospelov:2011ha,Billard:2014yka,Franco:2015pha,Schumann:2015cpa,Strigari:2016ztv,Dent:2016wcr,Chen:2016eab,Cerdeno:2016sfi,Dutta:2019oaj,Lang:2016zhv,Bertuzzo:2017tuf,Dutta:2017nht,Leyton:2017tza,AristizabalSierra:2017joc,Boehm:2018sux,Bell:2019egg,Newstead:2018muu,DARWIN:2020bnc,LZ:2021xov}.

The Sun is an ever-present source of natural neutrinos up to an energy of $18.8$ MeV. These will generate a background of recoil events in upcoming multi-ton-scale dark matter experiments~\cite{Akerib:2018lyp,Aprile:2017aty,Aalseth:2017fik,Cao:2014jsa,Agnese:2016cpb,Aalbers:2022dzr,Akerib:2022ort,Cooley:2022ufh}. The dominant channel of solar neutrino events in such experiments are electron recoils, where the large flux of $pp$ neutrinos generates $\sim$400 electron elastic scattering events (\vees) per ton-year of exposure~\cite{Schumann:2015cpa}. Because of this, experiments reaching $\mathcal{O}(10)$ ton-year exposures, like the next-generation dual-phase xenon experiment XLZD~\cite{Aalbers:2022dzr}, are even expected to be competitive with Borexino~\cite{Borexino:2008fkj,Bellini:2013lnn,Bellini:2014uqa,Agostini:2017cav,Borexino:2017uhp,Borexino:2017rsf,BOREXINO:2018ohr,BOREXINO:2017yyp,Agostini:2020mfq}, a 100-ton (fiducial mass) liquid scintillator that has led the measurement of solar neutrino fluxes for over a decade. Additionally, with low enough energy thresholds, nuclear recoils from coherent elastic neutrino-nucleus scattering (\cevns) of $^8$B neutrinos may also be present as a background in direct detection experiments---if so, this will constitute the first detection of \cevns with a natural neutrino source.

Although solar neutrinos are an exciting channel for novel physics searches, when it comes to performing these searches in dark-matter experiments, the situation is fraught. The reason is the existence of a large region of dark-matter parameter space called the ``neutrino fog''~\cite{OHare:2021utq} where a conclusive identification of a dark matter signal is rendered almost impossible due to the background of astrophysical neutrinos that also generate nuclear recoils, see e.g.~Refs.~\cite{Billard:2013qya, OHare:2016pjy, Dent:2016iht, Dent:2016wor, Gelmini:2018ogy, Gonzalez-Garcia:2018dep, Evans:2018bqy, Papoulias:2018uzy, Essig:2018tss, Wyenberg:2018eyv, Nikolic:2020fom, Munoz:2021sad, Calabrese:2021zfq, Sierra:2021axk, Carew:2023qrj, Tang:2023xub, Herrera:2023xun} for the many previous studies and variations on this idea. To circumvent this existential problem, recoil-based experiments would require extra information beyond just recoil energy to discriminate between the dark matter and neutrino events. 

Previous work has shown that only directional information is able to efficiently push through the neutrino fog~\cite{Grothaus:2014hja, O'Hare:2015mda, Mayet:2016zxu, OHare:2017rag, OHare:2020lva}. So far, directional detection of low-energy recoil tracks has only been shown to be experimentally feasible for time-projection chambers (TPCs) using gaseous targets~\cite{Battat:2016pap, Vahsen:2021gnb}. While gas targets have many limitations, for probing into the neutrino fog, a cost-balanced trade-off is possible using the high-resolution capabilities of highly-segmented micro-pattern gas detectors (MPGDs) which enable sub-mm spatial resolution and high signal-to-noise reconstructions of fully 3-dimensional ionisation tracks~\cite{Surrow:2022ptn}. This type of directional detection is referred to as ``recoil imaging''~\cite{Vahsen:2021gnb, OHare:2022jnx} to distinguish it from other, less powerful types of directional detection that seek to only measure spatial projections of ionisation or to infer directionality indirectly via daily modulation signals. Recoil imaging can provide a large amount of event-level information, but the detrimental effects of diffusion necessitate short drift lengths, which in turn implies that such an experiment can only be achieved in a modular configuration. A modular design also makes the path towards a larger-scale experiment clearer---an array of small-scale atmospheric-pressure modules is a much less daunting prospect than the naive statement of required total gas volumes of $\sim$1000 m$^3$ would suggest. Working towards this ultimate vision of a large-scale ``recoil observatory'' is the goal of the \Cygnus Consortium~\cite{Vahsen:2020pzb}---a global R\&D project that is currently tackling many of the discrete experimental challenges associated with constructing large-scale recoil imaging experiments. A competitive final experiment that enters the neutrino fog is thought to be achievable with 1000 m$^3$ of combined fiducial volume. Reaching this scale through a modular configuration naturally implies that there will be several stepping-stone detectors---these experiments, while smaller in scale, will already have sufficient target mass to obtain sensitivity to neutrinos via electron scattering. 

While astrophysical neutrino fluxes have enjoyed decades of dedicated study by experiments much larger in scale than are proposed for the first stages of \Cygnus, there are several notable advantages that \emph{directional} measurements could offer. One clear advantage is pointing---the reconstruction of angular distributions is beneficial for any sources of neutrinos with uncertain directionality, e.g. galactic supernovae. However, neutrino sources with \textit{known} directional distributions, in particular those originating from a single direction like the Sun or a neutrino beam, could also benefit from directional detection. Since there is a fixed kinematic relationship between a particle's initial energy and the energy and scattering angle of the subsequent recoil, in principle, a directional measurement allows event-by-event reconstruction of the particle's initial spectrum.

In many searches for neutrinos, this reconstruction is not possible~\cite{Bellini:2013lnn}, and is especially challenging at low energies. Recently the concept \textit{Theia}~\cite{Theia:2019non} was put forward, which proposes to enhance existing liquid scintillators with cutting-edge photon detectors to simultaneously detect Cherenkov light alongside scintillation. This signal---which is now utilised by Borexino~\cite{BOREXINO:2021efb, BOREXINO:2021xzc,BOREXINO:2023ygs}---enables directional sensitivity at lower energies than has been possible so far in the field. The concept we propose here would enable significantly lower energy and directionality thresholds for both nuclear and electron recoils, and could potentially enable a more precise and \textit{direct} measurement of the incoming neutrino direction. It could also be implemented in more compact experiments and tested at neutrino beams~\cite{AristizabalSierra:2021uob}.

The case for modern MPGD-based experiments to be neutrino detectors has been pointed out several times~\cite{Vahsen:2021gnb,OHare:2022jnx}, but it has yet to be quantified---this is the goal of the present article. This work complements ongoing gas simulation and detector R\&D within the \Cygnus~\cite{Vahsen:2020pzb,Schueler:2022lvr,Ghrear:2024rku} and CYGNO~\cite{Baracchini:2020btb,Amaro:2023dxb,Almeida:2023tgn,CYGNO:2023gud} collaborations. Rather than evaluate the prospects for a fixed detector configuration (which has so far only been optimised for a dark-matter nuclear recoil search~\cite{Vahsen:2020pzb}), we instead aim to determine the performance \emph{goals} that \Cygnus would need to reach in order to be a competitive neutrino experiment\footnote{Because of this generality, although we centre our discussion on gas detectors, our results can also be interpreted as requirements for the angular performance of other proposed directional technologies, such as those using solid state targets~\cite{Agafonova:2017ajg,Marshall:2020azl,Bernabei:2003ct,Belli:2020hit}, or other more radical designs~\cite{Drukier:2012hj,Capparelli:2014lua,OHare:2021cgj}.}. To this end, we will evaluate the prospects for both a final-stage low-density 1000 m$^3$ scale experiment that will be an ultimate dark matter and neutrino facility, as well as smaller stepping-stone experiments of $\sim$10 m$^3$ size. One of the virtues of gas-based experiments is the flexibility to choose and change the target medium and density, and so our results will directly inform the various possible operating configurations for these intermediate experiments.

Beginning in Sec.~\ref{sec:scattering}, we introduce the theoretical description of \cevns and \vees, highlighting their direction dependence. Then, in Sec.~\ref{sec:fluxes}, we apply these theoretical calculations to the solar neutrino fluxes and describe the resulting angular signatures. In Sec.~\ref{sec:TPCs}, we briefly summarise the current status of TPC technology and describe gas and simulation results that inform our definitions of several realistic detector performance benchmarks. Then, in Sec.~\ref{sec:nureconstruction}, we present our results on the sensitivity of gas TPCs to solar neutrinos, giving some recommendations and goals for future simulations and experimental validation. We conclude in Sec.~\ref{sec:conc}.

\section{Neutrino scattering}\label{sec:scattering}

\subsection{Elastic scattering kinematics}
\begin{figure*}
\begin{center}
\includegraphics[trim = 0mm 0 0mm 0mm, clip, width=0.99\textwidth]{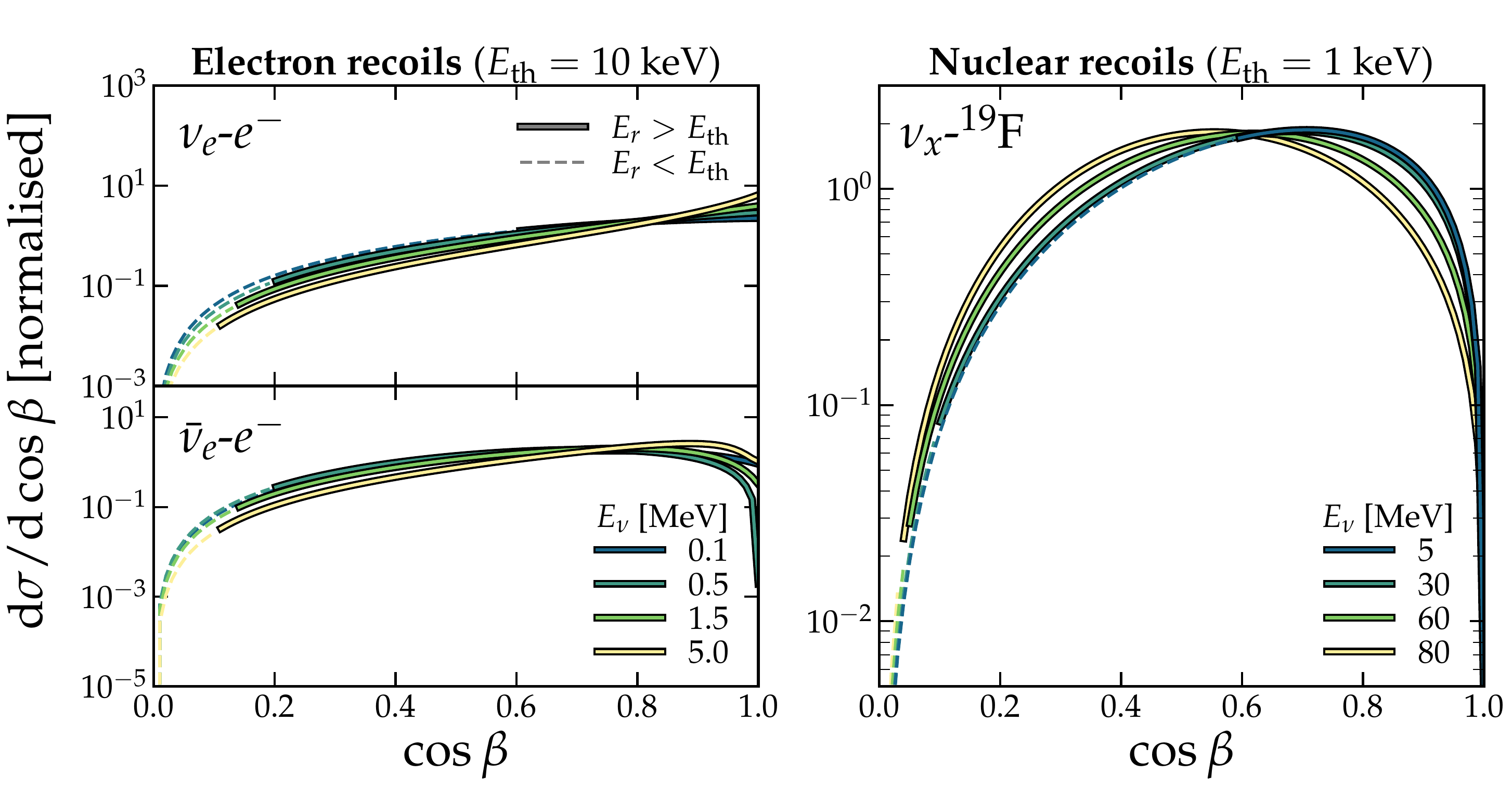}
\caption{Angular cross sections (normalised to one) as a function of the lab-frame neutrino scattering angle, $\beta$, between the neutrino direction and recoiling electrons (left) and fluorine nuclei (right). The different colours correspond to different neutrino energies at scales relevant for this study. The solid portion of each line corresponds to recoil energies scattering above 10 keV (left) and 1 \keVr (right). The two left-hand panels show the differences between the angular cross sections for $\nu_e$ and $\bar{\nu}_e$ for comparison, although only the former is relevant for this study.} 
\label{fig:AngularRecoilSpectra}
\end{center}
\end{figure*}

A directional gas TPC is designed to be able to detect and distinguish both electron and nuclear recoils via the differences in their track shape and ionisation profiles, at the same time as measuring their directions. For completeness and for reference in future studies, we will discuss the sensitivity of \Cygnus to both electron recoils via \vees, as well as nuclear recoils via \cevns. That said, it is worth anticipating the fact that the detectable event rate in a 10--1000 m$^3$-scale experiment will be dominated by electron recoils. For that reason, the sensitivity study we perform in the final section will focus primarily towards the measurements facilitated by \vees.

Before writing down specific interaction cross sections for \vees and \cevns, we can inspect the kinematics of elastic scattering common to both, and derive a general formula for the (nuclear/electron) recoil event rate as a function of recoil energy and direction. The kinematic expression for the angle, $\beta \in [0,\pi/2]$, between the neutrino direction, $\hat{{\bf q}}_\nu$, and the recoil direction, $\hat{{\bf q}}_r$ is~\cite{Vogel:1989iv},
\begin{equation}\label{eq:kinematics}
 \cos{\beta} = \hat{{\bf q}}_r \cdot \hat{{\bf q}}_\nu = \frac{E_\nu + m}{E_\nu}\sqrt{\frac{E_r}{E_r+2 m}} \, ,
\end{equation}
where $m$ will be either the nucleus mass ($m_N$) or electron mass ($m_e$). If we have some differential cross section $\textrm{d}\sigma/\textrm{d}E_r$, as a function of recoil energy, $E_r$, we can extend it to include the dependence on the direction of the recoiling target particle, $\Omega_r$, by first noting that the scattering has azimuthal symmetry about the incoming neutrino direction, i.e.~$\drm\Omega_\nu~=~2\pi \,\drm\cos\beta$. We can impose this relation with a delta function to get the direction-dependent cross section,
\begin{equation}\label{eq:doublecrosssection}
  \frac{\drm^2 \sigma}{\drm E_r \drm \Omega_r} = \dbd{ \sigma}{E_r} \, \frac{1}{2 \pi}\, \delta\left(\cos\beta - \frac{E_\nu + m}{E_\nu} \sqrt{\frac{E_r}{2 m}}\right) \,.
\end{equation}
With this cross section, we can write down the event rate per unit detector mass, as a function of the recoil energy, direction and time by multiplying it by the neutrino flux $\Phi$ and integrating over neutrino energies,
\begin{equation}\label{eq:nu_directionalrate}
  \frac{\drm^2 R_\nu(t)}{\drm E_r \drm\Omega_r} =  N_T \int_{E^{\rm min}_\nu} \frac{\drm^2 \sigma}{\drm E_r \drm \Omega_r}\frac{\drm^2 \Phi(t)}{\drm E_\nu \drm\Omega_\nu} \drm E_\nu \drm\Omega_\nu \, ,
\end{equation}
where $N_T$ is the number of target particles (i.e.~nuclei or electrons) per unit detector mass. We integrate only over kinematically permitted values of $E_\nu>E^{\rm min}_\nu (E_r)$, where,
\begin{equation}
    E_\nu^{\rm min}(E_r) = \sqrt{\frac{m^2 E_r}{E_r+2m}} \, ,
\end{equation}
is the minimum neutrino energy that can create a recoil with energy $E_r$.

We are currently interested in fluxes of neutrinos from a single source direction, i.e.~the Sun, although neutrinos from supernovae~\cite{Beacom:1998fj,Ando:2001zi,Tomas:2003xn,Abe:2016waf} or a neutrino beam~\cite{Abdullah:2020iiv,AristizabalSierra:2021uob} could also be described with the same treatment. We can therefore proceed by assuming the flux is a delta function towards a direction $\qhat_\odot$\footnote{We do not consider the $<0.25^\circ$ angular size of the solar core or the differing radial profiles of solar neutrino emission, although this is a subtlety that could perhaps be explored with an extremely large number of events~\cite{Davis:2016hil}.},
\begin{equation}
\frac{\mathrm{d}^{2} \Phi}{\mathrm{d} E_{\nu} \mathrm{d} \Omega_{\nu}}= \frac{\mathrm{d} \Phi}{\mathrm{d} E_{\nu}}\,\delta\left(\hat{\mathbf{q}}_{\nu}-\hat{\mathbf{q}}_{\odot}\right) \, ,
\end{equation}
where $\drm \Phi/\drm E_\nu$ is the neutrino energy flux. Substituting this into Eq.(\ref{eq:nu_directionalrate}) and manipulating the delta function leads to a relatively simple formula that effectively just enforces the kinematic relationship between $\beta$, $E_\nu$ and $E_r$,
\begin{equation}\label{eq:directionaleventrate}
  \frac{\drm^2 R_\nu}{\drm E_r \drm \Omega_r} =  \frac{N_T}{2\pi} \frac{\mathcal{E}^2}{ E_\nu^\textrm{min}} \left(\dbd{\sigma}{E_r}  \dbd{\Phi}{E_\nu} \right)\bigg|_{E_\nu = \mathcal{E}} \, ,
\end{equation}
for $\cos^{-1}(\qhat_r\cdot\qhat_\odot)<\pi/2$, and 0 otherwise. We also define,
\begin{equation}\label{eq:Eps}
  \mathcal{E} = \frac{m E_\nu^{\rm min}}{m (\hat{{\bf q}}_r \cdot \hat{{\bf q}}_\odot) - E_\nu^{\rm min}} \, .
\end{equation}
We can also write down the maximum recoil energy $E_r^{\rm max}$, that corresponds to $\beta = 0$,
\begin{equation}\label{eq:Emax}
E_r^{\rm max} = \frac{2E_\nu^2}{m + 2E_\nu} \,.
\end{equation}

\subsection{Coherent neutrino-nucleus scattering}
\cevns proceeds via a neutral current and is coherently enhanced at low momentum transfer by a factor that approximately scales with the number of neutrons squared~\cite{Freedman:1973yd,Freedman:1977, Drukier:1983gj}. At higher recoil energies $\gtrsim \mathcal{O}(10)$~\keV, the loss of coherence is described by the nuclear form factor $F(E_r)$, for which we use the standard Helm ansatz~\cite{Lewin:1995rx}. The differential \cevns cross section as a function of the nuclear recoil energy, $E_r$, and neutrino energy is given by~\cite{Freedman:1973yd,Freedman:1977,Drukier:1983gj}
\begin{equation}\label{eq:CEvNS}
 \dbd{\sigma_{\nu N}}{E_r}(E_r,E_\nu) = \frac{G_F^2}{4 \pi} Q^2_W m_N \left(1-\frac{m_N E_r}{2 E_\nu^2} \right) F^2(E_r) \, ,
\end{equation}
where $Q_W = A-Z - (1-4\sin^2\theta_W) Z$ is the weak hypercharge of a nucleus with mass number $A$ and atomic number $Z$, $G_F$ is the Fermi coupling constant, \mbox{$\sin^2{\theta_W} = 0.2387$} is the weak mixing angle~\cite{Erler:2004in}. Current measurements by COHERENT~\cite{Akimov:2017ade, COHERENT:2020iec, COHERENT:2020ybo} so far show consistency of the \cevns cross section with the Standard Model prediction. The incorporation of exotic mediators or non-standard interactions~\cite{Cerdeno:2016sfi,Bertuzzo:2017tuf,Boehm:2018sux,AristizabalSierra:2017joc} may lead to changes to the angular spectrum of recoils, but here we are focusing on the question of whether simply detecting the neutrino directions is feasible, so will leave an exploration of beyond-Standard-Model physics to future work.

The directional dependence of the \cevns cross sections for several specific neutrino energies is illustrated in the right-hand panel of Fig.~\ref{fig:AngularRecoilSpectra}. We see that the typical scattering angles are relatively wide for \cevns, compared to \vees (see next section) which are more focused in the forward direction. The effect of the finite nuclear recoil threshold (chosen here to be 1 keV) is to cut recoils scattering with the widest angles away from the Sun, as can be seen by the dashed segments of the lines in Fig.~\ref{fig:AngularRecoilSpectra}.

\subsection{Neutrino-electron elastic scattering}
While \cevns has so far only been detected with an artificial neutrino source, \vees has long been an essential channel for many neutrino experiments. Like \cevns, it is threshold-free and the direction of the outgoing electron is well-correlated with the initial neutrino direction, making it especially useful for low-energy neutrino fluxes. The \vees cross-section incorporates both charged and neutral current interactions, which also means that the event rate is higher by almost an order of magnitude for $\nu_e$ compared to other flavours. For neutrino flavours $i = e,\mu,\tau$, the \vees cross section is given by~\cite{Bahcall:1995mm},
\begin{equation}
\begin{aligned}
\frac{\drm \sigma_{\nu_i e}}{\drm E_r} =\frac{G_{F}^{2} m_{e}}{2 \pi} & \bigg[(g_{v}+x+g_{a})^{2} \\
&+(g_{v}+x-g_{a})^{2}\bigg(1-\frac{E_r}{E_{\nu}}\bigg)^{2} \\
&+(g_{a}^{2}-\left(g_{v}+x\right)^{2}) \frac{m_{e} E_r}{E_{\nu}^{2}}\bigg]\\
&+\frac{\pi \alpha^{2} \mu_{\nu}^{2}}{m_{e}^{2}}\bigg(\frac{1}{E_r}-\frac{1}{E_{\nu}}\bigg),
\end{aligned}
\end{equation}
where,
\begin{equation}
    g_{v}=\left\{\begin{array}{ll}
2 \sin ^{2} \theta_{W}+\frac{1}{2}, & \text { for } \nu_{e} \\
2 \sin ^{2} \theta_{W}-\frac{1}{2}, & \text { for } \nu_{\mu}, \nu_{\tau} \, ,
\end{array}\right.
\end{equation}
\begin{equation}
    g_{a}=\left\{\begin{array}{ll}
+\frac{1}{2}, & \text { for } \nu_{e} \\
-\frac{1}{2}, & \text { for } \nu_{\mu}, \nu_{\tau} \, ,
\end{array}\right.
\end{equation}
\begin{equation}
    x=\frac{\sqrt{2} \pi \alpha\left\langle r^{2}\right\rangle}{3 G_{F}} \, ,
\end{equation}
and for $\bar{\nu}$ contributions we must take $g_a \rightarrow -g_a$. In the above, $\langle r^{2}\rangle$ is the neutrino charge radius and $\mu_\nu$ is the neutrino magnetic moment, which can both be set to zero in accordance with current experimental data, but could also be accounted for and constrained. Electron events from \vees are typically much higher in energy than \cevns recoils for the same neutrino energy, with a maximum $E_r$ that can reach the MeV range for the most energetic solar fluxes. We show the angular cross section for \vees for specific neutrino energies around 1 MeV, and for both $\nu_e$ and $\bar{\nu}_e$ in the left hand panels of Fig.~\ref{fig:AngularRecoilSpectra} (the latter are shown for reference only, we do not study any fluxes of antineutrinos here). The \vees cross section is more strongly pointed in the forward direction than the \cevns cross section for the sub-MeV energy range dominated by the highest solar neutrino fluxes.


\section{Solar neutrinos}\label{sec:fluxes}
\begin{figure*}
\begin{center}
\includegraphics[trim = 0mm 0 0mm 0mm, clip, width=0.9\textwidth]{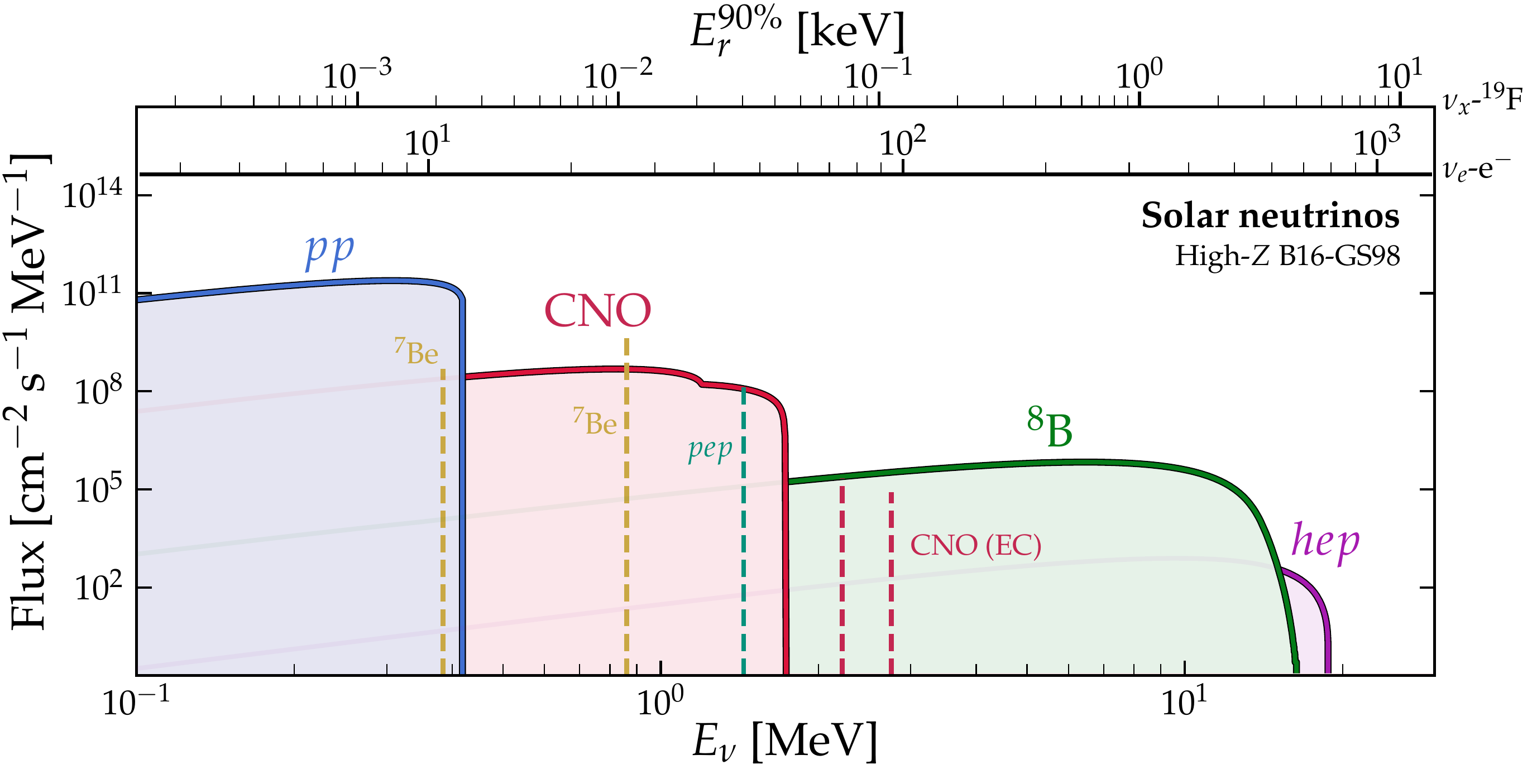}
\caption{Solar neutrino fluxes normalised to the B16-GS98 high metallicity Standard Solar Model~\cite{Vinyoles:2016djt}. The auxiliary axes show the recoil energies (for electrons and fluorine nuclei) below which 90\% of the events scatter for an equivalent value of $E_\nu$ on the lower axis. We colour each region by which source of neutrino is dominant at a given energy. Monoenergetic neutrino fluxes are shown by dashed lines.} 
\label{fig:NeutrinoFluxes}
\end{center}
\end{figure*}

\subsection{Fluxes}
For both nuclear and electron recoil channels, the dominant natural source of neutrino events will be the Sun. The Sun produces several well-understood fluxes of neutrinos from a variety of processes involved in its nuclear fusion cycles (see e.g.~Ref.~\cite{Gann:2021ndb} for a recent review). The fluxes are displayed in Fig.~\ref{fig:NeutrinoFluxes}. We also show $E^{90\%}_r$, which we define to be the recoil energy below which 90\% of all events fall. This is to give a sense of the required thresholds needed to detect the flux at a given neutrino energy. We show two axes, one for electron recoils and the other for $^{19}$F recoils.

There are two main sets of nuclear fusion reactions that are relevant for solar neutrinos: the PP chain and the CNO cycle. The latter is of little importance for the Sun's energy production but is of great importance for studying its composition. Fluxes of $\nu_e$ can either be produced in a continuous spectrum in a $\beta^+$ decay, or as a single line if a nuclide captures an electron.

The PP chain can be broken into five branches that each generate distinct fluxes of neutrinos. The PP-I branch involves the fusion of two protons to make helium via deuterium and leads to a very large flux of ``$pp$" neutrinos below 0.3 MeV that makes up around 99\% of the total emission. The Sun's luminosity in neutrinos can be closely linked to its luminosity in photons meaning we can infer the $pp$ flux normalisation to the sub-percent level. Some deuterium is also produced by the much less frequent proton-electron-proton ($pep$) reaction, a product of which is a monoenergetic flux of neutrinos at 1.44 MeV. 

The next branch, PP-II, fuses helium to make lithium, with an intermediate step involving electron capture by $^7$Be which generates two more lines at 0.862 and 0.384~MeV, with a 90:10 branching ratio. The nuclear energy is not sufficient to undergo $\beta^+$ decay, so there is no continuum flux associated with $^7$Be. The PPP-III chain involves the decay of $^8$B, generating a sizeable flux at much higher energies. This flux comes with an additional theoretical uncertainty relating to the energy of the excited $^8$Be$^*$ state. However, thanks in part to their relatively high energies, $^8$B neutrinos have been observed by several experiments~\cite{Super-Kamiokande:2001ljr,Borexino:2008fkj,Borexino:2017uhp,Super-Kamiokande:2010tar,KamLAND:2011fld,SNO:2011hxd,SNO:2018fch}, leading to a well-defined $\sim$2\% measurement of the flux. The final branch of the PP cycle is the helium-proton ($hep$) reaction which generates the highest energy solar neutrinos up to almost 20 MeV, however its tiny flux will likely remain unobserved, possibly until next-generation experiments such as DUNE~\cite{Abi:2020evt,Kelly:2019itm}, Hyper-Kamiokande~\cite{Abe:2018uyc}, JUNO~\cite{An:2015jdp,Djurcic:2015vqa} and the Jinping Neutrino Experiment~\cite{JinpingNeutrinoExperimentgroup:2016nol} begin operation.

The full CNO cycle has several branches, but only two (CNO-I and CNO-II) occur in the Sun. Within these branches, three continuous fluxes of neutrinos are generated from the $\beta^+$ decays of $^{13}$N, $^{15}$O, and $^{17}$F. These fluxes are very small and generate recoil spectra that are hidden underneath the $^7$Be and $pep$ lines. In Fig.~\ref{fig:NeutrinoFluxes} we also show the three often neglected CNO lines. These are present at energies around $2m_e$ higher than the CNO endpoints and originate from electron capture processes on the same three nuclei~\cite{Stonehill:2003zf,Villante:2014txa} (two of these lines are at almost exactly the same energy). These lines lead to a negligible event rate in all examples we study here (as will be seen in Fig.~\ref{fig:NeutrinoRates}). Only in the last few years have Borexino extracted the CNO neutrinos from the background and have measured their combined flux to be~\cite{Agostini:2020mfq,Borexino:2022pvu,BOREXINO:2023ygs},
\begin{equation}
    \Phi_{\rm CNO}= 6.7^{+1.2}_{-0.8} \times 10^8 \,{\rm cm}^{-2} \,{\rm s}^{-1}\, .
\end{equation}

All theoretical computations of solar neutrino fluxes are computed using a Standard Solar Model (SSM)---a simplified description of the Sun that can be calibrated against a number of observables. All of the solar model properties of the Sun are known to great accuracy, with the exception of its chemical composition, meaning the CNO fluxes, which are tied to the solar composition, are the most uncertain. The discussion surrounding the so-called solar abundance problem is typically framed in the context of the two models mentioned above which are referred to as high- and low-metallicity (Z), with a crucial distinguishing factor being the predicted fluxes of $^8$B and CNO neutrinos. An additional handle on the Sun's metal content via a firm measurement of the CNO neutrino flux would therefore be invaluable for resolving the solar abundance problem. Current Borexino measurements suggest a slight statistical preference for the high-metallicity models~\cite{Borexino:2022pvu}. We will not attempt to determine the scale of a \Cygnus-like experiment needed to resolve the abundance problem further, since it is likely to be prohibitive on its own, however it will be interesting to determine the scale and angular performance required to observe the flux. For the purposes of this explorative study, we normalise our solar neutrino fluxes to the ``B16'' high-Z SSM from Ref.~\cite{Vinyoles:2016djt}. See also the discussion of the various fluxes in Ref.~\cite{Vitagliano:2019yzm}.

\subsection{Scattering rates}
Continuing from Eq.(\ref{eq:directionaleventrate}), we can incorporate some additional corrections to the scattering rate specific to solar neutrinos. The full direction and time dependence can be written as,
\begin{align}\label{eq:solarnu}
  \frac{\drm^2 R(t)}{\drm E_r \drm \Omega_r} &=  \sum_{j={e,\mu,\tau}} \frac{P_{ej}(E_\nu)}{2\pi m_N}\left[ 1 + 2 e \cos\left(\frac{2\pi(t- t_\nu)}{T_\nu}\right) \right] \nonumber \\
   &\times\frac{\mathcal{E}(t)^2}{ E_\nu^\textrm{min}} \left(\dbd{\sigma_j}{E_r}  \dbd{\Phi}{E_\nu} \right)\bigg|_{E_\nu = \mathcal{E}(t)} \, ,
\end{align}
which accounts for neutrino oscillations from electron neutrinos into neutrino flavour $j$ with probability $P_{ej}$, as well as the annual variation ($T_{\nu} = 1$~year) in the Earth-Sun distance due to the eccentricity of the Earth's orbit ($e = 0.016722$). The latter leads to a $\pm$3.4\% seasonal modulation of the flux with a maximum occurring on $t_{\nu} = 3$rd January (see e.g.~measurements made by Refs.~\cite{Super-Kamiokande:2001ljr,BOREXINO:2017yyp,BOREXINO:2022wuy}).

Nuclear fusion processes in the Sun emit only $\nu_e$, however these neutrinos experience a strong energy-dependent matter oscillations inside the Sun as well as vacuum oscillations between the Sun and Earth. The neutrinos arriving at Earth can, therefore, no longer be treated in their flavour eigenstate. To calculate the rate, we must sum over the three neutrino species weighted by $P_{ei}$, the probability for the electron neutrino to have oscillated into flavour $i$ when it is detected. Recall that \cevns is flavour-blind\footnote{Up to radiative corrections to the cross-section which do introduce flavour-dependence~\cite{Mishra:2023jlq}.}, so the cross section $\textrm{d}\sigma_j/\textrm{d}E_r$ is the same for all $j$, meaning the sum is only necessary for electron rates.

We adopt the MSW-LMA solution for neutrino oscillations~\cite{Mikheyev:1985zog,Wolfenstein:1977ue}. In the vacuum dominated regime at low neutrino energies, the electron-neutrino survival probability can be expressed as (see e.g.~\cite{Friedland:2000cp,Antonelli:2012qu,Vissani:2017dto} for further discussion),
\begin{equation}
    P_{e e}=\cos ^{4} \theta_{13}\left(\frac{1}{2}+\frac{1}{2} \cos 2 \theta_{\odot} \cos 2 \theta_{12}\right)+\sin ^{4} \theta_{13} \, ,
\end{equation}
where,
\begin{equation}
\cos 2 \theta_{\odot}=\frac{\cos 2 \theta_{12}-\xi_{\odot}}{\left(1-2 \xi_{\odot} \cos 2 \theta_{12}+\xi_{\odot}^{2}\right)^{1 / 2}}
\end{equation}
is the mixing angle inside the Sun, and,
\begin{equation}
   \begin{aligned}
\xi_{\odot} \equiv \frac{l_{\nu}}{l_{0}} &=\frac{2 \sqrt{2} G_{F} \rho_{\odot} Y_{e} \cos ^{2} \theta_{13}}{m_{n}} \frac{E}{\Delta m_{12}^{2}} \\
&=0.208\times \cos ^{2} \theta_{13}\left(\frac{E_\nu}{1 \mathrm{MeV}}\right) \\
&\times\left(\frac{\Delta m_{12}^2}{7.32\times10^{-5}\,\text{eV}^2}\right)\left(\frac{\rho_{\odot} Y_{e}}{100 \mathrm{~g} \mathrm{~cm}^{-3}}\right),
\end{aligned}
\end{equation}
is the ratio of the oscillation length in vacuum and the refraction length in matter. We take a standard averaged value for the solar matter density, $\rho_\odot$, and number of electrons per nucleon $Y_e$. By ignoring matter oscillations inside the Earth, we are implicitly assuming observation during the day. A day/night asymmetry due to this effect makes the electron-neutrino flux slightly larger during the night~\cite{Lisi:1997yc}. However, this is only at the 3\% level for high solar neutrino energies where the \vees rates are significantly lower. We take $\Delta m^2_{12} = 7.32\times10^{-5}$~eV$^2$, $\theta_{12} \approx 33^\circ$ and $\theta_{13} \approx 0.15^\circ$ from the three-neutrino global fit of Ref.~\cite{Esteban:2018azc}. The energy dependence of $P_{ee}(E_\nu)$ will be slightly different for each neutrino flux since the radial profiles of their emission are not identical, however, this effect is very small and can be safely neglected here for better than 1\% precision.

The MSW-LMA solution predicts $P_{ee}$ is around 0.55 at $pp$ energies and then decreases to around 0.3 by the endpoint of the $^8$B spectrum. The neutrino oscillation probabilities are also constrained experimentally at $pp$, $pep$, $^7$Be, and $^8$B energies by Borexino~\cite{BOREXINO:2018ohr,Borexino:2017uhp} and SNO~\cite{SNO:2011hxd} in general agreement with the MSW-LMA solution. There is a gap in the measured $P_{ee}$ between 2--10~MeV where it is relatively unconstrained.

\subsection{Recoil spectra and event numbers}

\begin{figure*}
\begin{center}
\includegraphics[trim = 0mm 0 0mm 0mm, clip, width=0.9\textwidth]{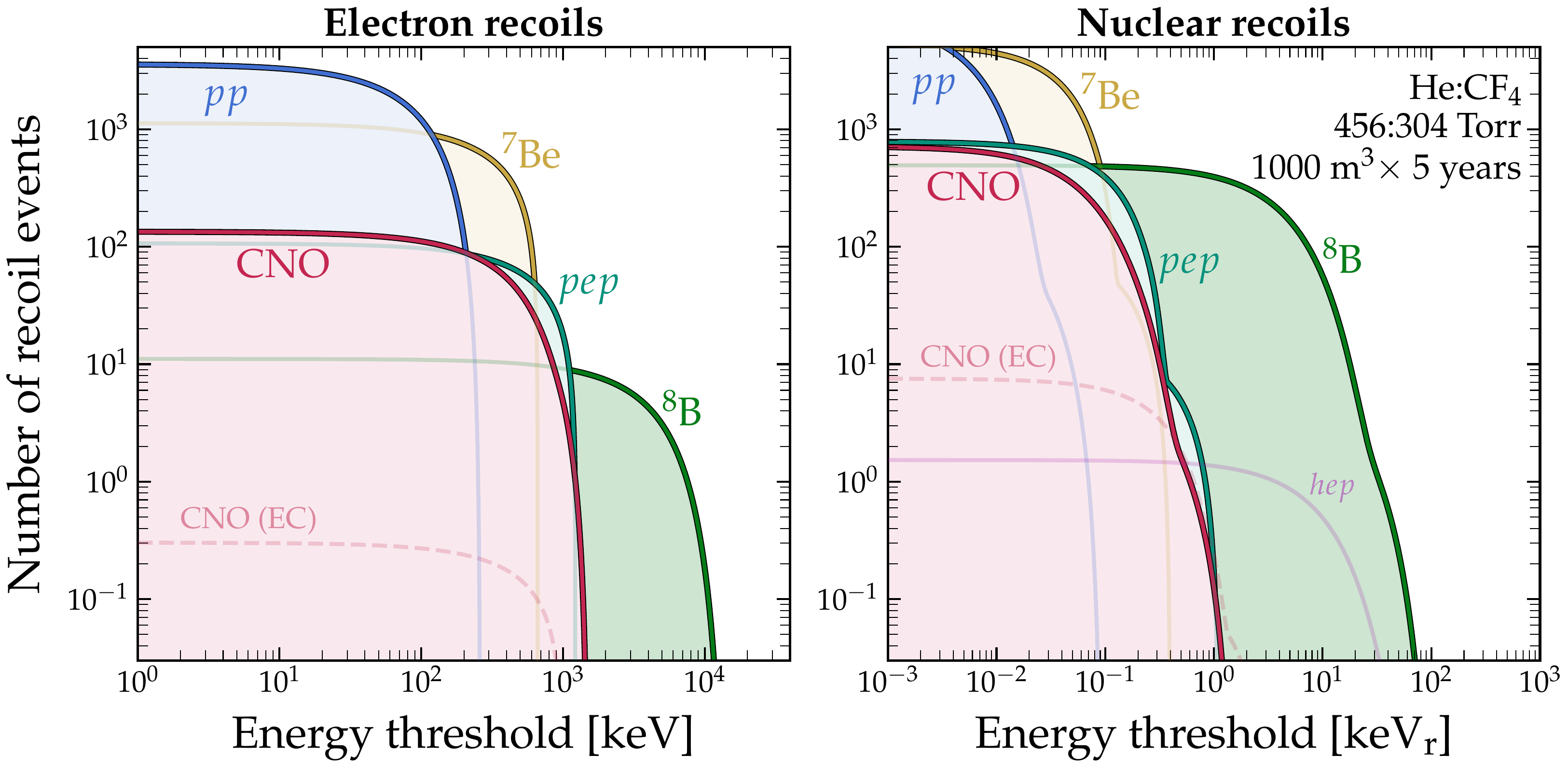}
\caption{Expected number of recoil events as a function of the energy threshold above which $\textrm{d}R/\textrm{d}E_r$ is integrated. We show the rates for electron events on the left, and nuclear recoils on the right, both assuming a baseline 60:40 He:CF$_4$ electron-drift gas mixture at 1 atmosphere (456:304 Torr) that we will adopt for the final sensitivity analysis. We assume a volume of 1000 m$^3$ and a total data-taking time of 5 years.} 
\label{fig:NeutrinoRates}
\end{center}
\end{figure*}

\begin{figure*}
\begin{center}
\includegraphics[trim = 0mm 0 0mm 0mm, clip, width=0.9\textwidth]{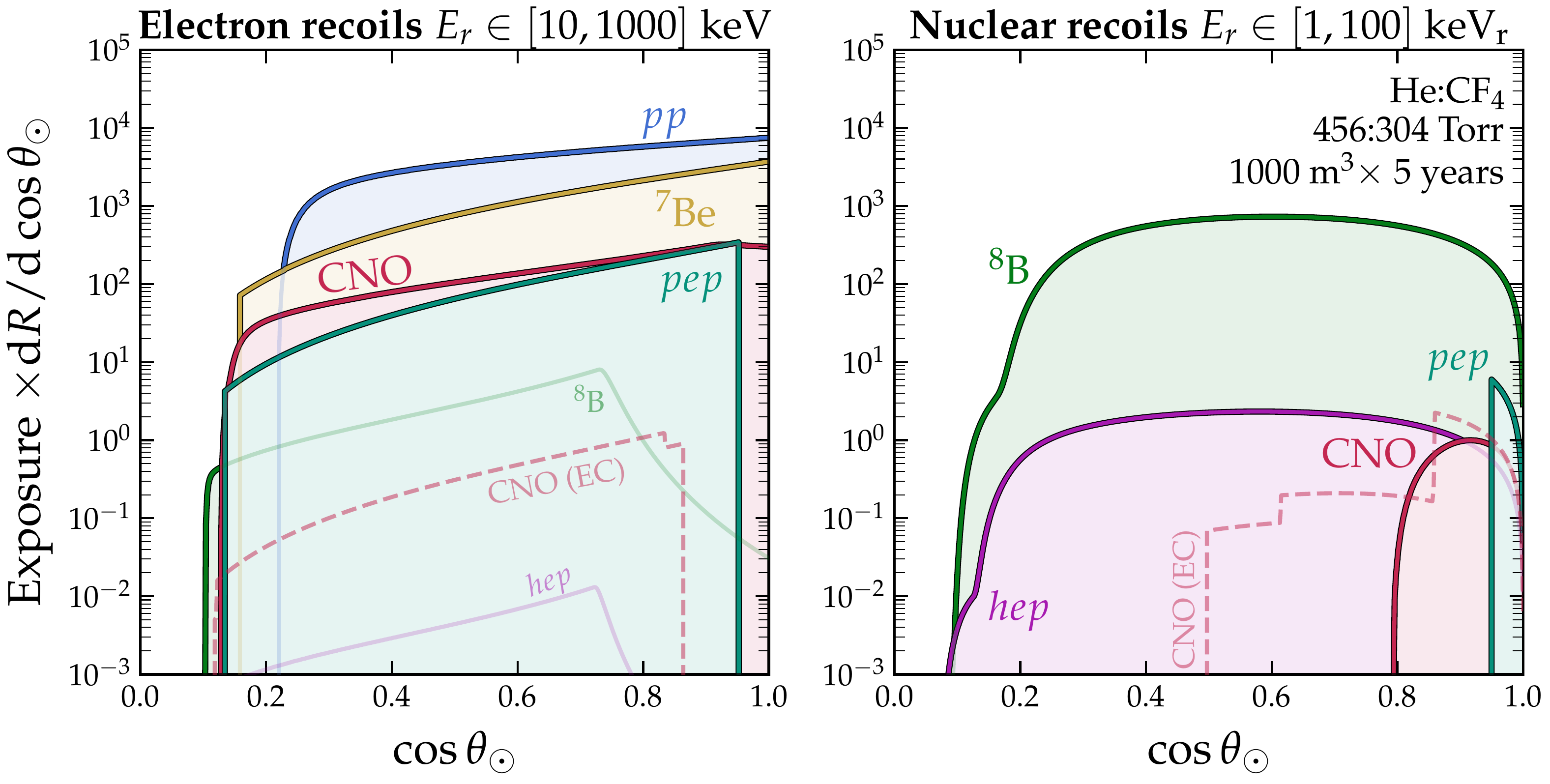}
\caption{Expected number of recoil events as a function of the angle away from the Sun, $\cos{\theta_\odot}$. We show the rates for electron events on the left, and nuclear recoils on the right, both assuming the baseline 60:40 He:CF$_4$ electron-drift gas mixture at 1 atmosphere (456:304 Torr), as in Fig.~\ref{fig:NeutrinoRates}. We assume a volume of 1000 m$^3$ and a total data-taking time of 5 years.} 
\label{fig:NeutrinoRates_costh}
\end{center}
\end{figure*}

The measurement of low energy solar neutrinos via directional electron recoils is not a new idea. Largely-forgotten work from the 1990s~\cite{Seguinot:1992zu,Arzarello:1994jv,Arpesella:1996uc}, proposed the use of a TPC filled with high densities of gases like He and CF$_4$ to detect solar-neutrino electron recoils with energies $\gtrsim$100 keV. While most fluxes generating high numbers of electron recoils are now much better understood, the neutrino flux spectroscopy made possible through event-by-event directionality is still an intriguing prospect. A modern gas TPC with a 1000 m$^3$ volume at atmospheric pressure or higher could make \emph{directional} measurements down to $\mathcal{O}(10)$~keV energies, much lower than the $\sim$160 keV threshold of Borexino.

Modern gas TPCs are also expected to have excellent electron/nuclear discrimination, hence we wish to incorporate both sources of events in this study. In Fig.~\ref{fig:NeutrinoRates} we show the recoil spectra as a function of both recoil energy (upper panels) as well as $\cos{\theta_\odot}$ which is the recoil's apparent arrival angle away from the Sun. We show the spectra for both electron recoils (left) and nuclear recoils (right).

Throughout, we will set 1 \keVr and 10 keV as hard lower limits on the recoil rate of nuclear and electron recoils respectively. Most nuclear recoils due to \cevns will be from $^8$B neutrinos, with a rate of around 77 events per ton-year above 1 keV assuming a $^{19}$F based gas. A helium target, due to its lighter mass, has a lower cross section, but can provide many more events at higher energies. For instance, the maximum $E_r$ from $^8$B neutrinos is 30~\keVr for $^{19}$F, but 142~\keVr for $^4$He.

Electrons, on the other hand, have much lower cross sections with neutrinos, but the momentum transfers are significantly larger, leading to higher energy recoils. This entails the majority of the electron event rate being from $pp$ neutrinos, with around 432 events per ton-year of He, between 10 keV and the maximum energy of 264 keV. The second and third most dominant sources are $^7$Be and $pep$ which respectively generate 145 and 13 events per ton-year of He above 10 keV. CNO neutrinos would generate events up to around 1500 keV, with a mean energy of around 200 keV and a total rate of around 17 events per ton-year of He above 10 keV when the three fluxes are combined.

As for the angular spectra, we see that both are focused in the forward half of the sky, but with the nuclear recoil scattering angles typically being much larger at around $80^\circ$. This would lead to a distinctive ring-like pattern around the Sun if viewed projected on the sky. The electron events are peaked towards $\cos{\theta_\odot} = 1$, with the only exception in this case being the $^8$B flux which is suppressed at small scattering angles because we have cut off the high energy part of the spectrum above 1000 keV. Track lengths inside the gas are correlated with the recoil energy and we must be mindful of recoils which would leave tracks that would not be contained within the TPC volume and hence would be vetoed. This is not an issue for any of the nuclear recoil energies associated with solar neutrinos, however it becomes problematic for electrons higher than around an MeV, as we will discuss in the next section.

\begin{figure}
\begin{center}
\includegraphics[trim = 0mm 0 0mm 0mm, clip, width=0.49\textwidth]{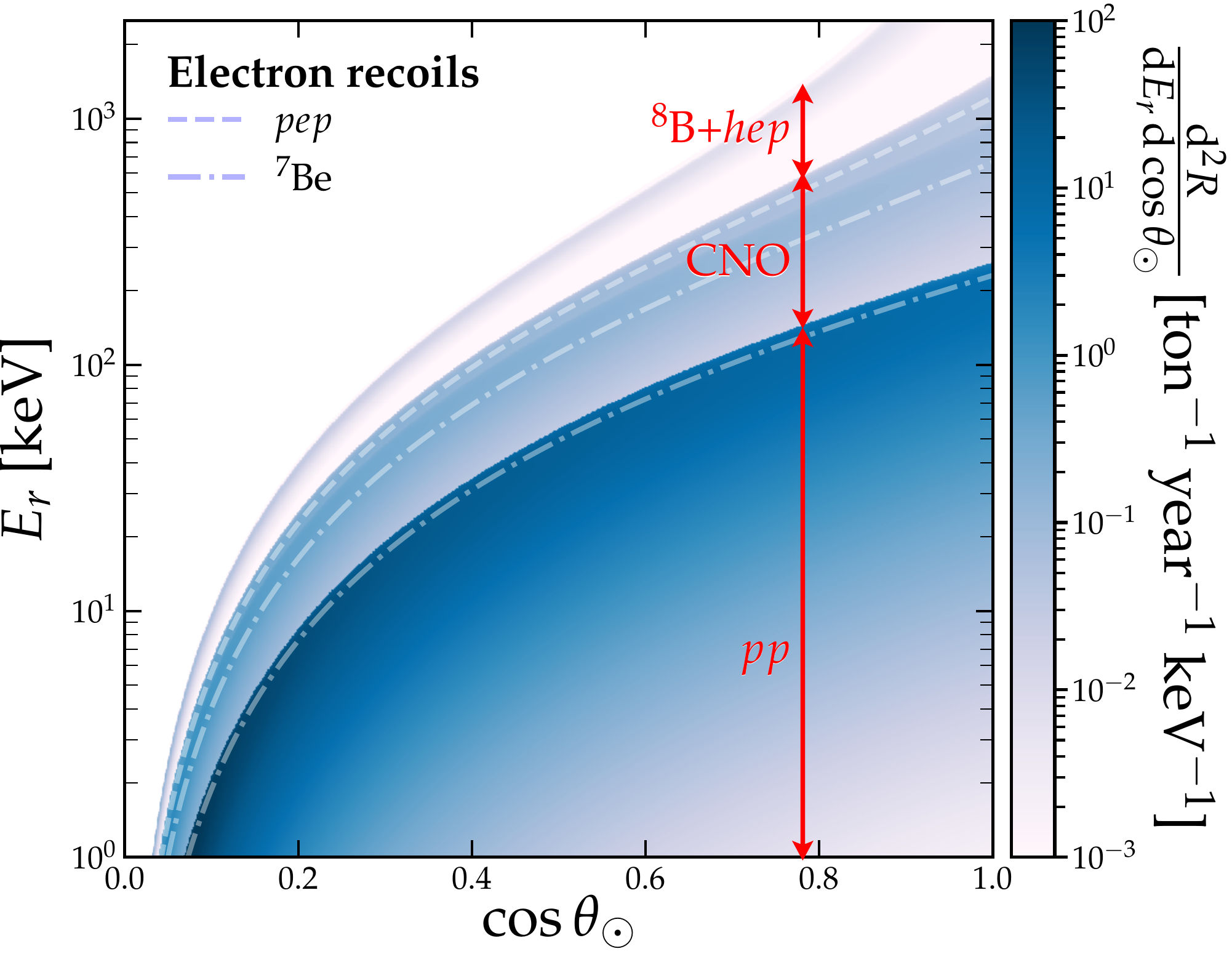}
\includegraphics[trim = 0mm 0 0mm 0mm, clip, width=0.49\textwidth]{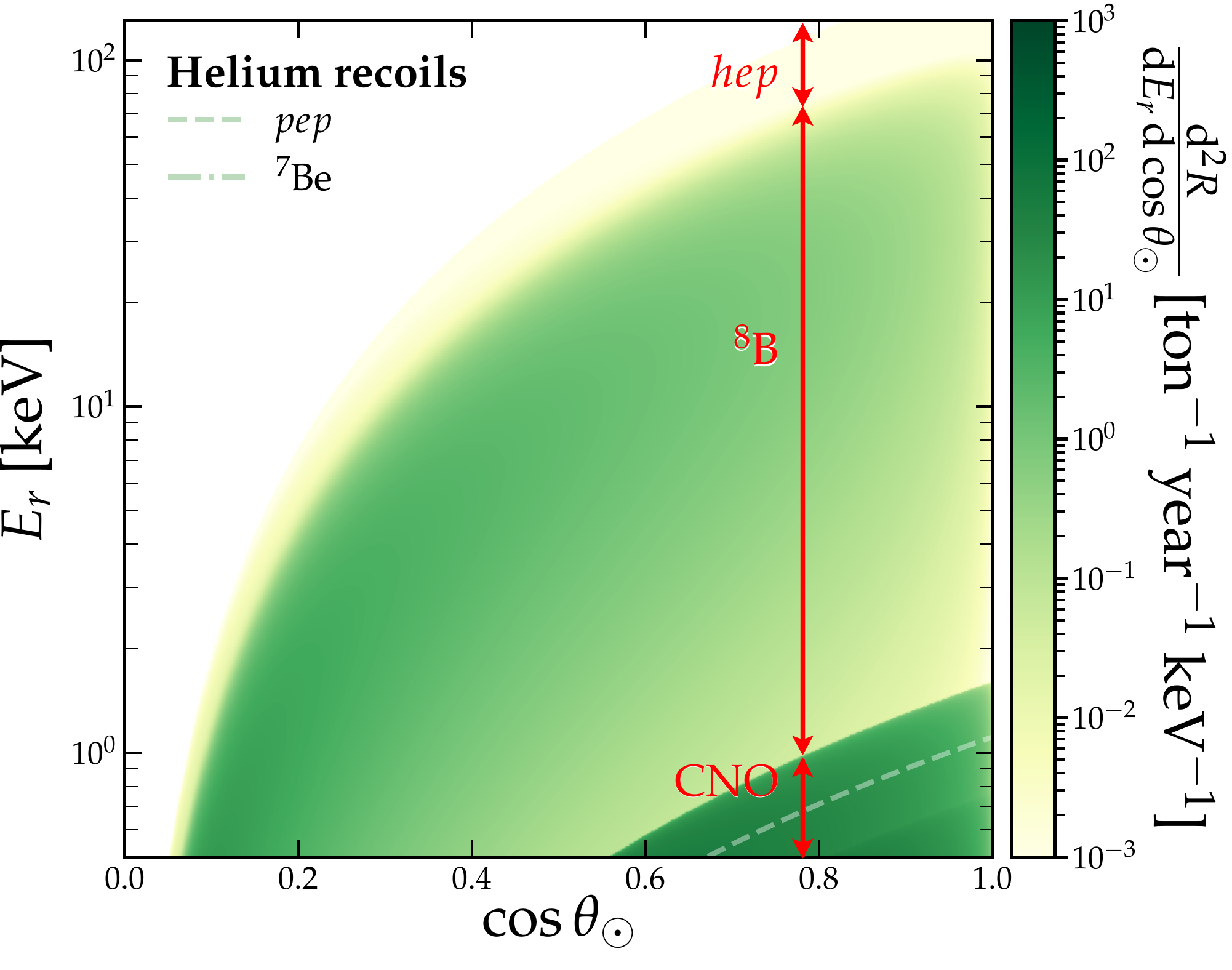}
\caption{Event rate versus recoil energy and angle with respect to the direction of the Sun. A value of $\cos \theta_\odot = 1$ corresponds to recoils aligning with the Sun. Since there is a one-to-one correspondence between recoil energy and angle, there are windows of this space in which CNO fluxes can be isolated. The lowest energy window shows primarily $pp$ neutrinos, whereas the highest energies are dominated by $^8$B and $hep$ neutrinos. The small window for CNO neutrinos, however, is also contaminated by the monoenergetic fluxes of $pep$ and $^7$Be neutrinos, which we show as dashed and dot-dashed lines respectively. This intermediate band of energies and angles would need to be targeted to make a measurement of the CNO flux.} 
\label{fig:SolarNu_Ercosth}
\end{center}
\end{figure}
Given the difficulty in measuring the elusive CNO flux in even state-of-the-art liquid scintillators, it may seem hopeless to attempt a directional measurement in gas. However, the most obvious novel aspect of directionality is in background rejection. When it comes to CNO neutrinos, the most problematic ``background'' will be the events coming from \emph{other} solar neutrinos (as can be seen in the overlapping of the CNO, $pep$ and $^7$Be electron event rates in Fig.~\ref{fig:NeutrinoRates}). Furthermore, as we hinted at in the introduction, directionality is novel in another way when dealing with a signal originating from a single direction. Given the known position of the Sun, the combined measurement of recoil energy and direction permits event-by-event reconstruction of the neutrino energy spectrum. 

To demonstrate this, Fig.~\ref{fig:SolarNu_Ercosth} shows the energy and angular distributions along a single plane. Each point in this 2-dimensional space corresponds to a \emph{unique} original neutrino energy and as such the distribution in this space maps directly back to the original neutrino energy spectrum. Contrast this fact with the upper panels of Fig.~\ref{fig:NeutrinoRates}, where a measured $E_r$ can correspond to a range of neutrino energies. To aid in interpreting Fig.~\ref{fig:SolarNu_Ercosth} we have overlaid some arrows indicating which parts of this 2-d space are occupied predominantly by different sources of neutrino. Since the $^7$Be and $pep$ fluxes are monoenergetic, they correspond to curved lines in this space which we have also overlaid. Notice that there are narrow windows in this space where the CNO rate is the largest contribution. The only region where this is the case when observing just $E_r$ is a small window between 1250 and 1500 keV.

Therefore, a measurement of a set of recoil directions and energies should allow a better combined measurement of these different fluxes, and with good performance, an empirical measurement of the solar neutrino energy spectrum. In practice, this will place a high demand on the energy resolution and tracking of the detector, but given what we have discussed so far, and the fact that solar neutrino electron recoils are not the primary target for \Cygnus, we can see that the science payoff from simply achieving excellent directional sensitivity is substantial.

\subsection{Empirically measuring the neutrino energy spectrum}
\begin{figure*}
\begin{center}
\includegraphics[trim = 0mm 0 0mm 0mm, clip, width=0.49\textwidth]{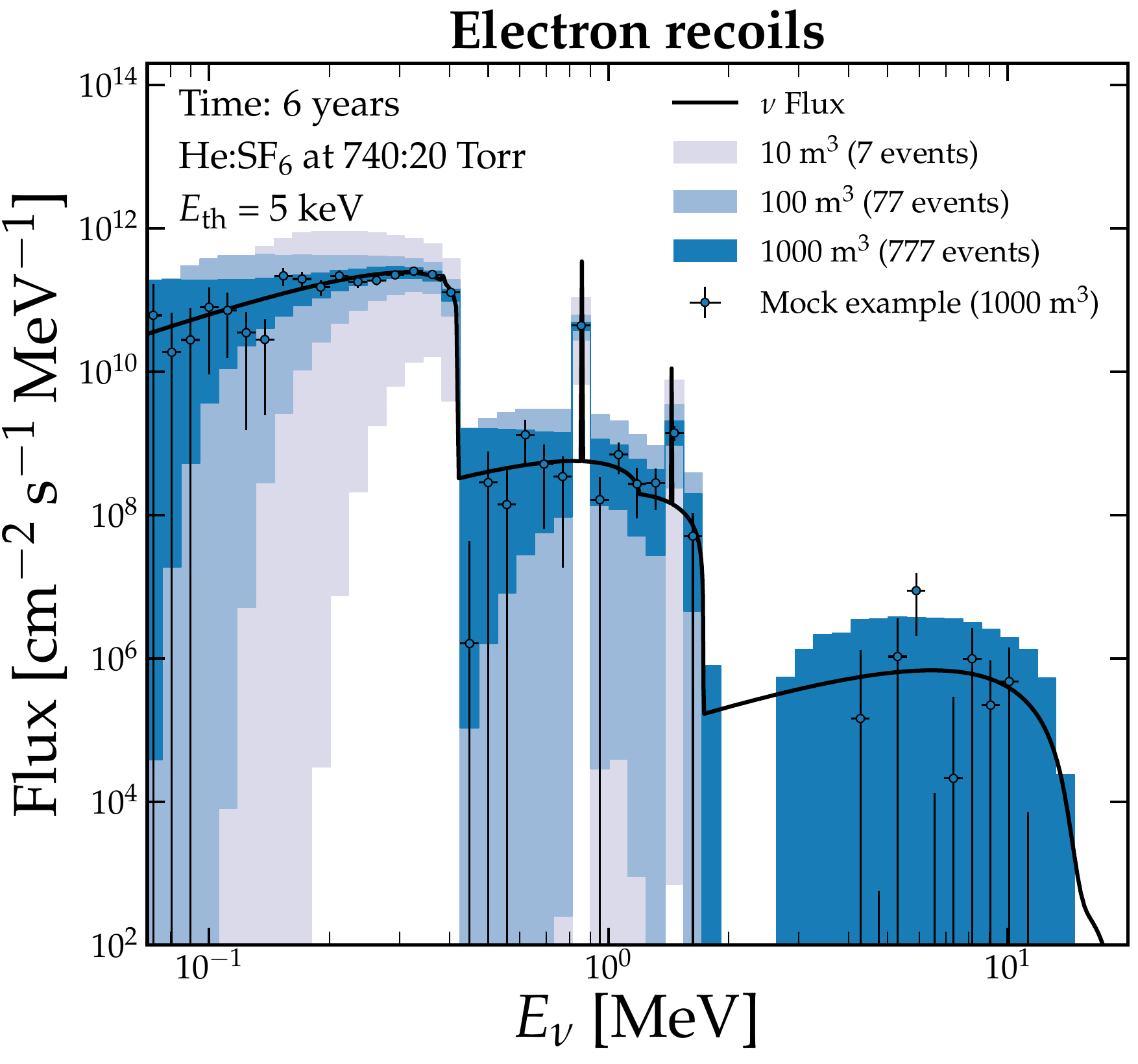}
\includegraphics[trim = 0mm 0 0mm 0mm, clip, width=0.49\textwidth]{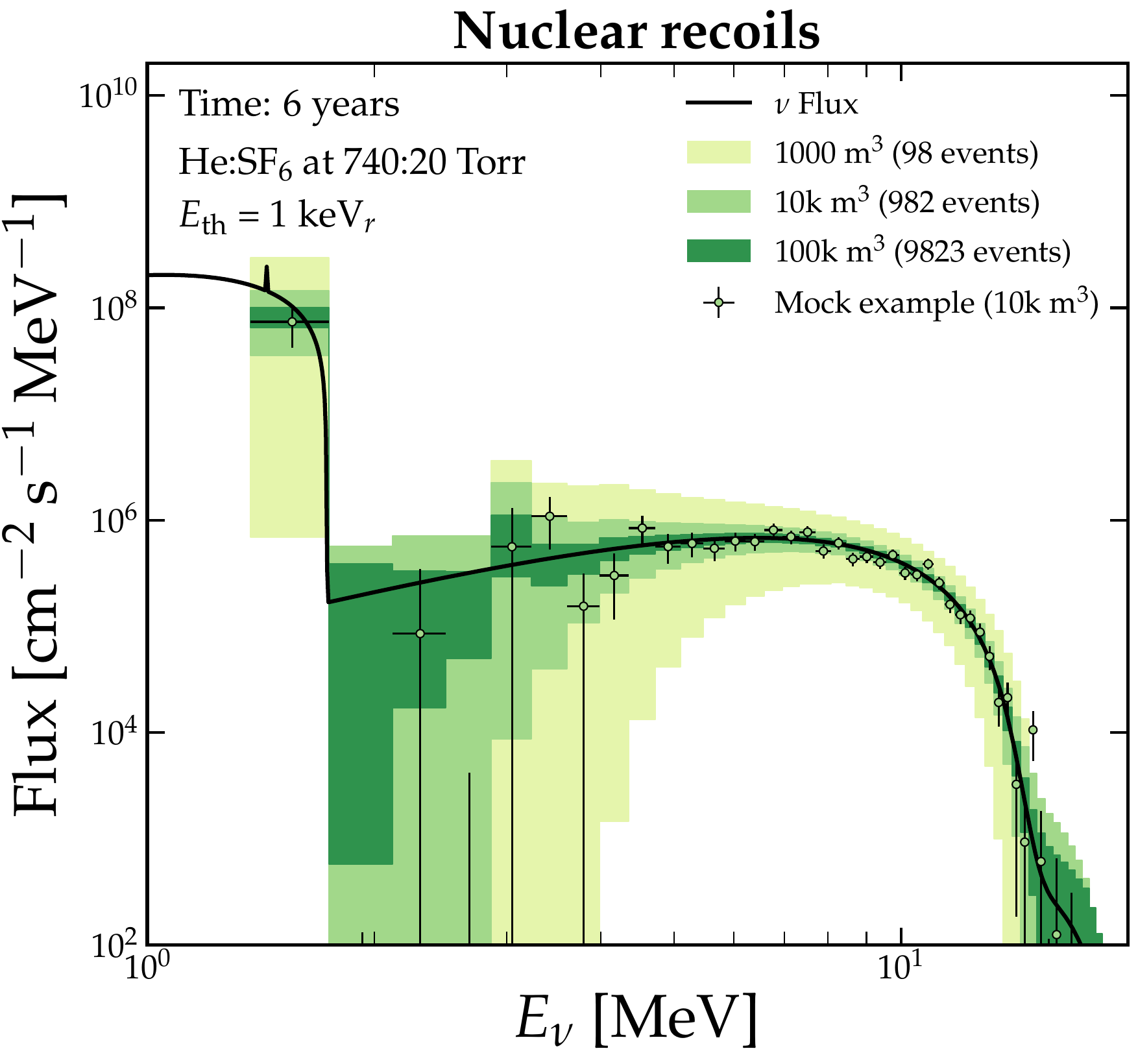}
\caption{Solar neutrino spectra reconstructed from \vees and \cevns event recoil energies and direction information alone. In this demonstrative example we assume that recoils are perfectly measured above 5 keV and 1 \keVr respectively. The shading from light to dark corresponds to increasing TPC volume, assuming a fixed gas ratio of 740:20 Torr He:SF$_6$ at 1 atmosphere. As expected, with more events, the reconstructed flux is improved. Each shaded region encloses 95\% of all reconstructed spectra collected over a large number of Monte Carlo pseudo-experiments.} 
\label{fig:NuFlux_reconstruction}
\end{center}
\end{figure*}

If we know $\qhat_0$, $E_r$ and $\qhat_r$, we can complete the kinematic relationship and derive $E_\nu$ event-by-event. The reconstructed distribution of $E_\nu$ will not be precisely $\drm\Phi/\drm E_\nu$, but will rather be weighted by the range of kinematically permitted values of $E_\nu$. Converting this distribution to the flux is then straightforwardly achieved by considering the rate of observed events as a function of neutrino energy,
\begin{equation}
    \dbd{\Phi}{E_\nu} = \dbd{R}{E_\nu} \bigg(N_T\int_{E_0}^{E_1} \dbd{\sigma}{E_r} \,  \drm E_r\bigg)^{-1} \, ,
\end{equation}
which can be determined for neutrino energies $E_\nu>E^{\rm min}_\nu(E_r^{\rm th})$. The limits of the integration are given by the range of recoil energies $E_r \in [E^{\rm th}_r,E^{\rm max}_r]$ that the experiment is sensitive to, but modified to not extend beyond the kinematically permitted values of $E_r$ as a function of $E_\nu$, which will be different for \cevns and \vees.

We first wish to get a sense of the required statistics to observe the fluxes over some range of neutrino energies, before we determine the effects of the performance of the experiment itself. As an illustrative example of the scope of this kind of measurement, we show some mock examples of reconstructed fluxes in Fig.~\ref{fig:NuFlux_reconstruction}. How tightly the fluxes are reconstructed depends upon how many events are observed and therefore the volume of the experiment. We assume here a He and SF$_6$ mixture at 740:20 Torr (1 atmosphere). This is slightly different from our chosen baseline gas that we will introduce later, since we wish to illustrate both nuclear and electron recoils simultaneously. However, the only important number here is the number of events which we have displayed on each panel. We see that with around 10 electron events, only the low energy $pp$ flux is accessible, but by around 1000 events even the flux out to $^8$B energies is accessible. For nuclear recoils, however, the $^8$B flux could measured, but only beginning for around 10--50 events, needing a 1000 m$^3$ volume or larger. The gas mixture assumed here is a majority helium at atmospheric pressure---with more SF$_6$, or alternatively CF$_4$, these volumes could be reduced, at the cost of some directional sensitivity.

\section{Directional gas TPCs}\label{sec:TPCs}
\begin{figure}
\begin{center}
\includegraphics[trim = 0mm 0 0mm 0mm, clip, width=0.45\textwidth]{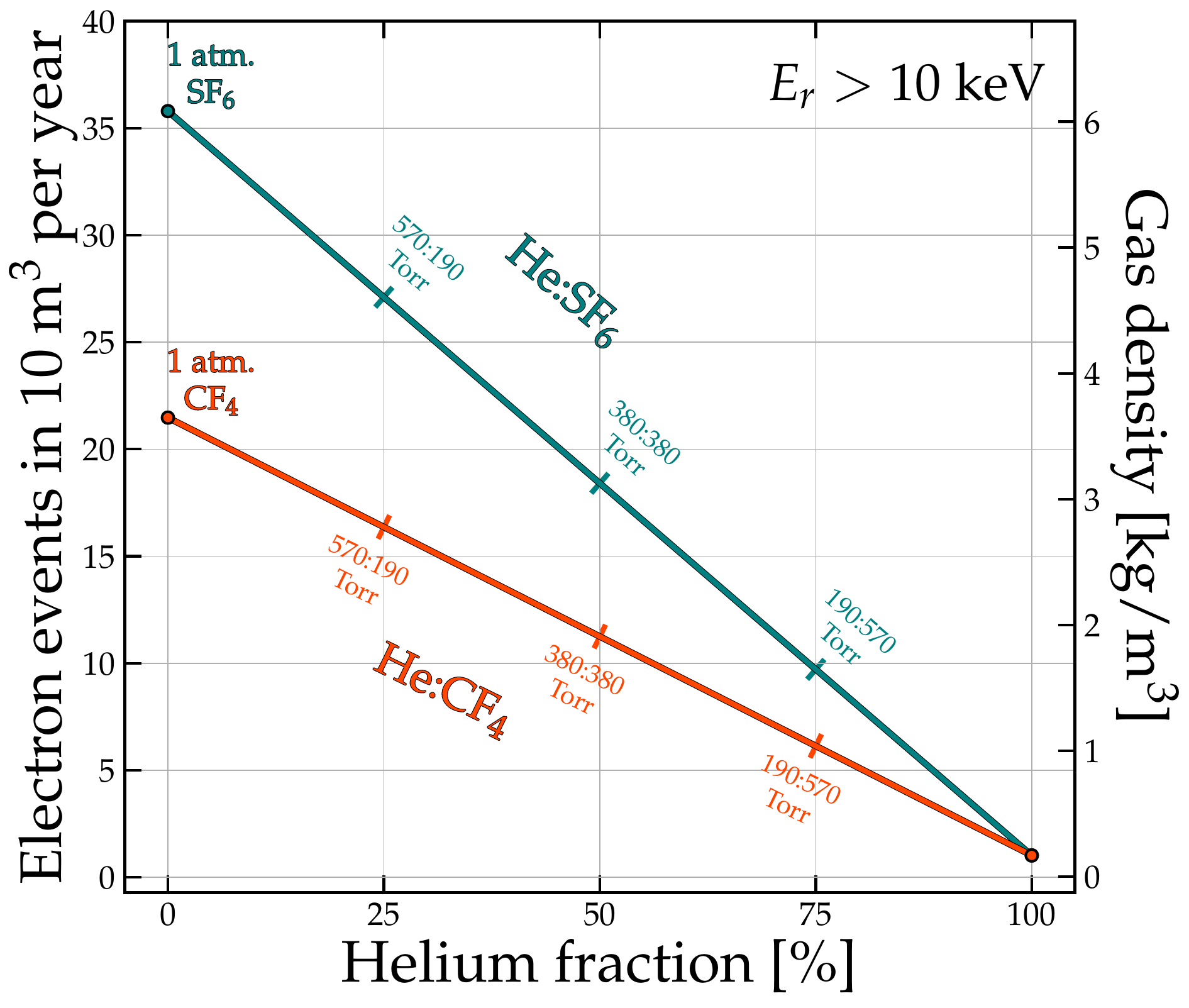}
\caption{Expected solar-neutrino-induced electron recoils above 10 keV in a 10-m$^3$ detector volume over one year plotted against the fraction of helium mixed with two potential gases, CF$_4$ and SF$_6$. The total pressure is fixed at one atmosphere. We also show the gas density corresponding to the mixture on the right. The lower gradient of the CF$_4$ line relative to the SF$_6$ line indicates that the latter gas has a better trade-off between electron count versus gas density when the pressure is kept at one atmosphere (even though the overall event numbers are smaller). A lower gas density improves directional performance, but would reduce the rate of \textit{nuclear} recoils.} 
\label{fig:GasDensity_vs_Events}
\end{center}
\end{figure}

Gas TPCs typically require low-density targets to obtain good directionality. Since most directional TPC proposals have been optimised exclusively to have sensitivity to dark-matter-induced nuclear recoils, some of our choices in operating parameters will be influenced by this bias. Some preferred fill gases include the fluorine-based CF$_4$ and SF$_6$, as well as He. For concreteness, we will fix our gas mixture to a 60\%:40\% mixture of He:CF$_4$ at atmospheric pressure, which has been studied recently in the context of the CYGNO experiment~\cite{Amaro:2023dxb,Almeida:2023tgn,CYGNO:2023gud}. CF$_4$ and SF$_6$ have almost identical values of $Z/A$, however when fixed to be at one atmosphere in-mixture with helium, CF$_4$ would provide a lower gas density and therefore better directionality. This is shown in Fig.~\ref{fig:GasDensity_vs_Events} which displays the number of expected electron events in a 10-m$^3$ module over one year for different mixtures where the total pressure is kept at one atmosphere. As discussed further in Ref.~\cite{Vahsen:2020pzb}, designing for an atmospheric pressure vessel would make the experiment significantly easier and cheaper to operate.

The low-pressure gas TPC is the most mature directional detection technology~\cite{Vahsen:2021gnb}. In this scheme, the recoils generate a track of ionisation in the gas volume, and an electric field transports the resulting charge to an amplification and readout plane. The full three dimensions of a recoil track can be reconstructed by combining the 2d measurement of the ionisation charge distribution at the readout plane, with the third dimension inferred by sampling the transported signal as a function of time. The projection of the track along this third dimension, parallel to the drift field, is found by multiplying the relative timing of the signal with the known drift velocity of the charge in the gas. In some designs, the ionisation electrons can be transported directly. In others, a target gas with high electron affinity is used where the ionisation electrons rapidly combine with surrounding gas molecules near the interaction site to form negative ions, which then drift to the readout plane. This latter method, so-called `negative ion drift' (NID), can help suppress the diffusion of the ionisation track to the thermal limit and thus preserve more of the track geometry prior to readout~\cite{SnowdenIfft:2013iy}. Recent work has explored NID operation at atmospheric pressure~\cite{Baracchini:2017ysg}.

NID gases are well known to improve the performance of dark-matter TPCs by limiting diffusion of the drift charge. Moreover, it has also been discovered that minority carriers in NID gases can enable fiducialisation in the drift direction. Fiducialisation refers to the ability to locate events within the detector volume, which is crucial for background rejection. Full 3d fiducialisation in NID gases was first demonstrated with CS$_2$~\cite{Snowden-Ifft:2014taa}, and more recently with SF$_6$~\cite{Phan:2016veo}. SF$_6$ has a number of attractive properties in the context of dark matter and has received substantial recent interest in the community~\cite{Ikeda:2017jvy, Baracchini:2017ysg, phdthorpe, Ikeda:2020pex}. 

Given the $\mathcal{O}({\rm keV})$ energies expected for dark-matter-induced nuclear recoils, and the typical energy required to create an electron-ion pair in gas targets, $W\approx20$\,eV, a recoil will produce $\mathcal{O}(10^2$--$10^3)$ primary ionisation electrons. To enhance this signal, a gas amplification device is used, consisting of a carefully designed region of high electric field where avalanche multiplication occurs. In some cases, this amplification device is integral to the readout---Multi-Wire Proportional Chambers (MWPCs) and Micromegas for instance---while in others, the gas amplification and readout are distinct, as with Gas Electron Multiplier (GEM) amplification followed by a separate charge readout via silicon pixel chips, or by an optical readout.

As for the choice of gas, we need to choose between a NID gas like SF$_6$ and the electron drift gas CF$_4$. One potential drawback of SF$_6$ is that it tends to result in approximately two orders of magnitude lower gas avalanche gains than is typical for commonly used electron drift gases. In detectors using electron drift, the substantially higher avalanche gain will generally lead to greatly improved charge sensitivity. With a high-resolution charge readout, 3d fiducialisation is also possible via different means~\cite{Lewis:2014poa}. Charge tends to diffuse more in electron drift detectors which reduces directionality at the $\mathcal{O}(1$--$10$ keV) energies needed to detect recoils from \cevns---this tips the balance towards favouring NID gases in that case, and for dark matter. On the other hand, for solar-neutrino-induced electron recoils, which are present out to much higher energies but whose tracks possess much lower ionisation densities, it is possible that electron drift could be the better option. This is one reason why we tend towards CF$_4$ in the present work because our sensitivity is driven predominantly by electron recoils. However, in the ultimate \Cygnus experiment, we would ideally like simultaneous sensitivity to low-energy nuclear recoils and electron recoils, so both options, as well as hybrids of the two, will continue to be explored.

\subsection{TPC performance}\label{sec:performance}
\begin{figure*}
\begin{center}
\includegraphics[trim = 0mm 0 0mm 0mm, clip, width=0.95\textwidth]{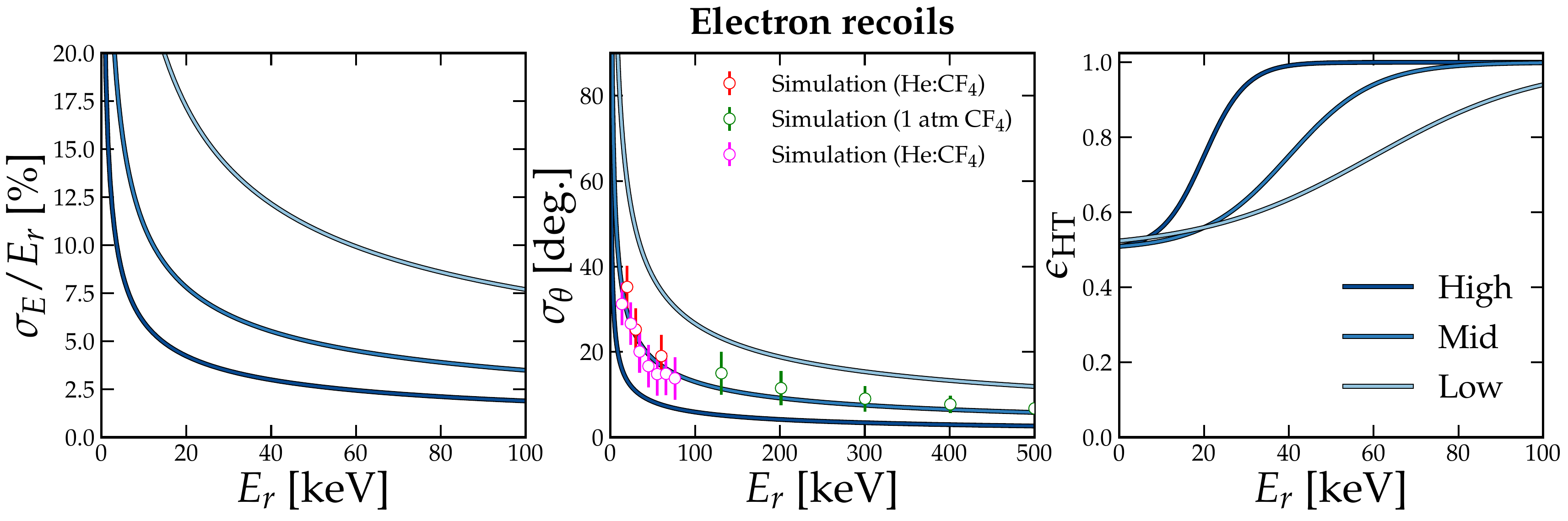}
\caption{Energy-dependent performance curves for electron recoils: energy resolution, axial angular resolution, and head-tail recognition efficiency. Each panel has three lines that correspond to the definitions of our three benchmark configurations: ``low'', ``mid'' and ``high'' coloured from light to dark blue. These three benchmarks should be interpreted as ``currently achievable'',  ``realistic with further optimisation'' and ``optimistic'' respectively. The middle benchmark is based on gas simulation and electron track reconstruction studies on similar setups to what we propose here (He:CF$_4$ mixtures at atmospheric pressure), as can be seen by the data points with statistical error bars taken from Ref.~\cite{SamueleThesis}.} 
\label{fig:PerformanceParameters_ERs}
\end{center}
\end{figure*}

\begin{figure*}
\begin{center}
\includegraphics[trim = 0mm 0 0mm 0mm, clip, width=0.99\textwidth]{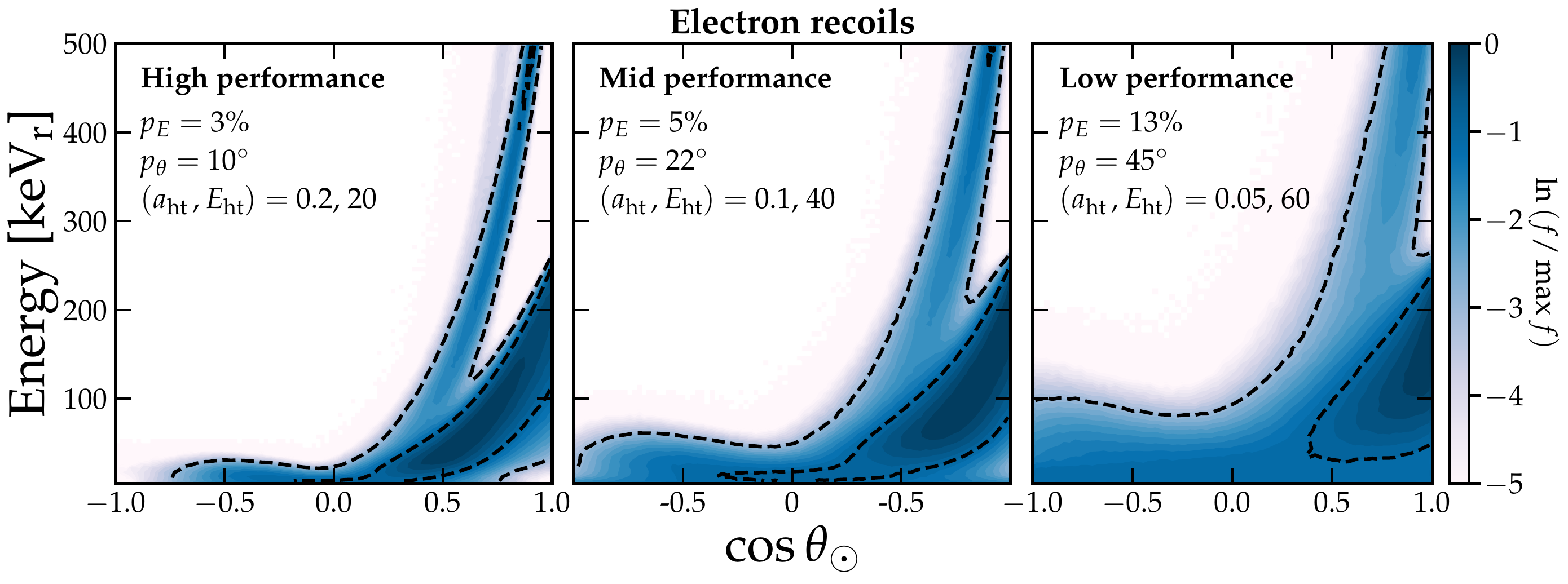}
\includegraphics[trim = 0mm 0 0mm 0mm, clip, width=0.99\textwidth]{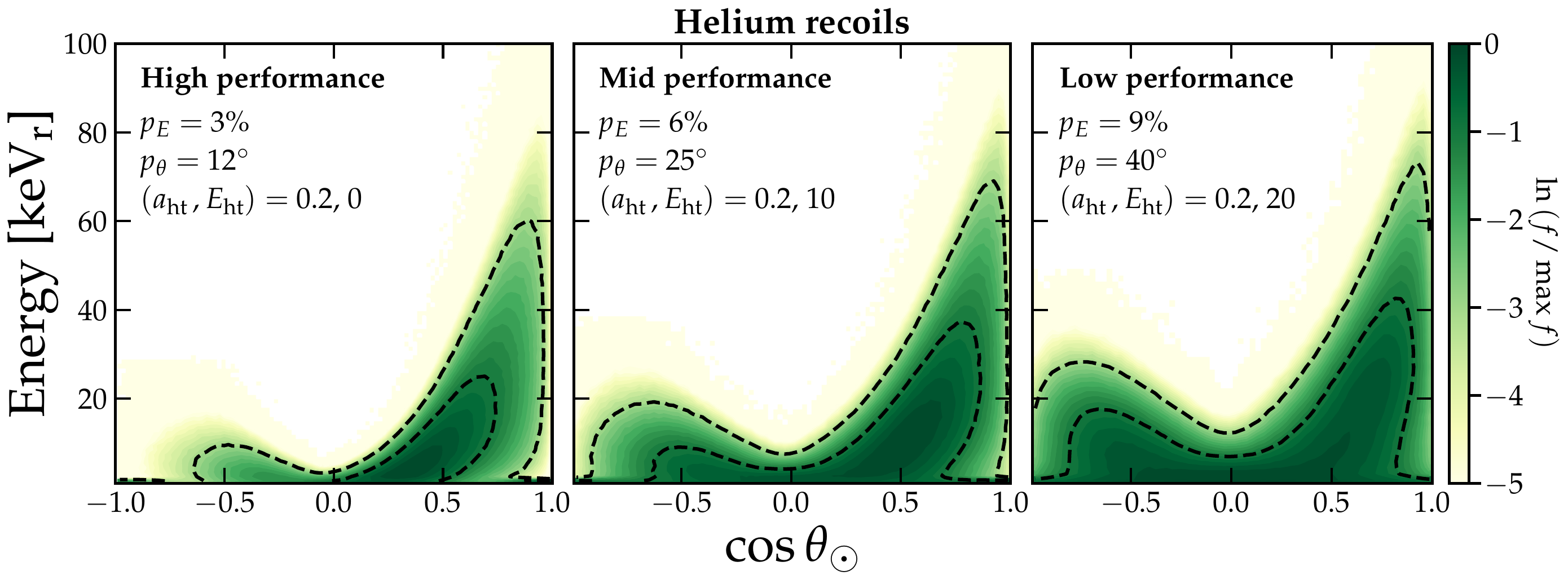}
\caption{Distributions of recoil energy and angle away from the Sun for electron recoils (top row) and helium recoils (bottom row). The probability distribution function in this space, $f$, is shown normalised by its maximum value and on a logarithmic colour scale. The three panels in each case correspond to the three performance benchmarks discussed in Sec.~\ref{sec:performance} that we form our later results around. They are ordered from high to low in terms of decreasing direction and energy reconstruction performance following the energy dependence shown in Fig.~\ref{fig:PerformanceParameters_ERs}. The labels shown in each panel ($p_E$, $p_\theta$, $a_{\rm HT}$, and $E_{\rm HT}$) refer to the parameters of the analytic performance curves defined in Sec.~\ref{sec:performance}. The dashed lines enclose 68\% and 95\% of the distributions in each case, and the colour scale encodes the logarithmic height of the distribution normalised by its maximum value.} 
\label{fig:SmearedDistributions}
\end{center}
\end{figure*}

Since we wish to use energy and angular information to reconstruct $E_\nu$, it is important to consider the performance of the TPC. The most important consideration will be how precisely the recoil energy and directions can be reconstructed using spatially resolved ionisation tracks. Our goal is not to estimate the performance of any specific readout technology here (as in Ref.~\cite{Vahsen:2020pzb} for instance), rather, we wish to set performance benchmarks that can be set as targets for ongoing simulation and experimental work. Nevertheless, in order to remain grounded in reality, we will explore a range of potential performance parameters that are inspired by ongoing simulations of gases that are of interest to \Cygnus~\cite{Vahsen:2020pzb} and CYGNO~\cite{Amaro:2023dxb,Almeida:2023tgn,CYGNO:2023gud}, as well as some optimistic projections for what could be achieved with additional optimisation and bespoke track reconstruction algorithms.

There are four distinct detector performance metrics that we incorporate: (1) angular resolution $\sigma_\theta$, (2) energy resolution $\sigma_E$, (3) head-tail recognition efficiency $\varepsilon_{\rm HT}$, and (4) event-level detection efficiency, $\varepsilon$. All of these are functions of recoil energy, and are different for nuclear and electron recoils. We do not incorporate the effects of nuclear quenching, so all of our nuclear recoil energies, including the quoted threshold energies, are treated as true recoil energies in units of \keVr. 

To incorporate these effects we perform a Monte Carlo simulation. The first step is to generate primary (true) neutrino energies and their recoil energies by drawing from the distribution,
\begin{equation}
    f(E_r,E_\nu) \propto \dbd{\sigma_T}{E_r}\dbd{\Phi}{E_\nu} \Theta(E_\nu - E_\nu^{\rm min}(E_r)) \, .
\end{equation}
where,
\begin{align}
    \text{Nuclear recoils}: &\quad \dbd{\sigma_T}{E_r} = \sum_{i=1}^{n_N}\zeta_i\dbd{\sigma_{N_i}}{E_r}, \nonumber \\
    \text{Electron recoils}: &\quad \dbd{\sigma_T}{E_r} = \sum_{i=e,\mu,\tau} P_{ei}(E_\nu)\dbd{\sigma_{e\nu_i}}{E_r} \, .
\end{align}
The $\zeta_i$ are the fractions of the total target mass comprised of nucleus $N_i$, for a target made of $n_N$ different kinds of nuclei. 

Then, for each pair of neutrino and recoil energy, we can calculate the scattering angle from Eq.(\ref{eq:kinematics}). We work in a solar coordinate system where the neutrino recoil directions peak towards $\cos{\theta_\odot} = 1$. In our Monte Carlo simulations, we also keep track of an azimuthal angle around this direction $\phi_\odot$, though we will assume that all distributions including the neutrino recoils and the backgrounds are flat in this direction. By assuming we can write all recoil vectors in this coordinate system, we are implicitly assuming that our TPC's readout can provide fully three-dimensional tracks. This can be achieved using MPGDs and pixel chips, as well as optical CCD or CMOS-based readouts if they are augmented with PMTs to provide timing information from which the third spatial dimension can be inferred.

We assume that the energy resolution at a given energy can be described by a Gaussian with width $\sigma_E$. This means that for each generated $E_r$ we draw another recoil energy $E^\prime_r$ from the distribution,
\begin{equation}
    K_E\left(E_{r}, E_{r}^{\prime}\right)=\frac{1}{\sqrt{2 \pi} \sigma_E\left(E_{r}\right)} \exp\left(-\frac{\left(E_{r}-E_{r}^{\prime}\right)^{2}}{2 \sigma_E^{2}\left(E_{r}\right)}\right) \, .
\end{equation}
We then simulate the effect of angular resolution by taking the initial recoil direction $\hat{\mathbf{q}} = (\sin{\theta_\odot}\cos{\phi_\odot},\sin{\theta_\odot}\sin{\phi_\odot},\cos{\theta_\odot})$ and sampling another recoil direction $\hat{\mathbf{q}}^\prime$ from a von Mises-Fisher distribution,
\begin{equation}
K_{\theta}(\hat{\mathbf{q}}, \hat{\mathbf{q}}^\prime,E_r)=\frac{\kappa(E_r)}{4 \pi \sinh \kappa(E_r)} \exp \left(\kappa(E_r) \mathbf{q} \cdot \mathbf{q}^\prime \right) \, ,
\end{equation}
where the parameter that specifies the variance of the von Mises-Fisher distribution is $\kappa$. Rather than quoting this parameter, we will express the angular resolution in terms of the root-mean-squared axial angular resolution $\sigma_\theta$, which is a monotonic function of $\kappa$ that can be evaluated numerically. The root-mean-squared angle between two axes approaches 1 radian for completely uncorrelated directions (i.e.~the limit of no directional sensitivity). The von Mises-Fisher distribution has the advantage over, for example, a truncated Gaussian, in that it reflects the fact that the distribution exists on the sphere.


Next, we implement the effect of an imperfect head-tail recognition (i.e.~correctly determining the vector sense of $\hat{\mathbf{q}}$) by taking each recoil and drawing a value of either \mbox{$+1$} or \mbox{$-1$} with a probability $\varepsilon_{\rm HT}(E_r)$ and then multiplying the recoil vector by that number. 

Finally, after performing all the above steps, we can discard recoils with a probability $\varepsilon(E_r)$ to simulate the effect of a finite event-level detection efficiency, which will suppress the event rate close to threshold. 

Putting these effects together, we can see that the recoil distribution can be written mathematically as,
\begin{equation}
\begin{aligned}
    \frac{\drm^2 R_{\rm obs}}{\drm E_r \drm \Omega_q} =& \int \varepsilon(E^\prime_r) \, K_E(E_r,E^\prime_r) \, K_{\theta}(\hat{\mathbf{q}}, \hat{\mathbf{q}}^\prime,E_r) \nonumber \\
    &\left[ \varepsilon_{\rm HT}(E^\prime_r)\frac{\drm^2 R}{\drm E^\prime_r \drm \Omega^\prime_q} +  \big(1-\varepsilon_{\rm HT}(E^\prime_r)\big) \frac{\drm^2 R}{\drm E^\prime_r \drm \Omega^\prime_{-q}} \right] \nonumber \\ \, &\quad \drm E^\prime_r \, \drm\Omega^\prime_q\, .
    \end{aligned}
\end{equation}


\subsubsection{Head-tail efficiency}
The head-tail recognition efficiency approaches 0.5 at low recoil energies where the ionisation track is either too small or generates too little ionisation to infer the direction with any accuracy. We parameterise the energy dependence with the following functional form,
\begin{equation}
    \varepsilon_{\rm HT}(E_r) = \frac{1}{2}\left(\frac{1}{1+e^{-a_{\rm HT}(E_r-E_{\rm HT}))}}+1 \right) \, ,
\end{equation}
where, for example $(a_{\rm HT},E_{\rm HT}) = 0.0065,  38.0)$ was found for He recoils in 755:5 Torr He:SF$_6$ ~\cite{Vahsen:2020pzb}.

In the case of nuclear recoils, the limited head-tail recognition is driven by the fact that most of this information is contained at a relatively limited level in the $\mathrm{d}E/\mathrm{d}x$ profile along the track, which is challenging to infer when the tracks are very small compared to the diffusion scale or the spatial resolution. In contrast, for electron recoils at the relevant energies, the head and the tail of the track are much more straightforwardly extracted. Unlike nuclear recoils, which are on the stopping side of the Bragg peak, the electrons deposit \textit{more} ionisation as they slow down. So the tail can be extracted easily from the part of the track that curls up and contains the most ionisation. That said, the fact that electrons follow highly non-linear trajectories through the gas, obtaining the correct \textit{initial} recoil direction is still challenging towards lower energies. In this case, our head-tail recognition efficiency is better thought of as a proxy for the probability of correctly assigning a direction to the head of the electron track. Like the nuclear recoil case, this will be effectively randomly assigned at the lowest energies. Our representative curves in Fig.~\ref{fig:PerformanceParameters_ERs} are inspired by recent simulations performed by the CYGNO collaboration on this issue~\cite{Amaro:2023dxb,Almeida:2023tgn,CYGNO:2023gud}.

\subsubsection{Angular resolution}

In Ref.~\cite{Vahsen:2020pzb}, a good fit to the energy dependence of the RMS axial angular resolution was found to be \mbox{$\sigma_\theta(E_r) = \frac{a}{\sqrt{\left(b+E_r^{c}\right)}}+d$}. We will attempt to simplify the discussion of our results by distilling this basic shape into a single parameter that can be interpreted straightforwardly. Instead, we will use,
\begin{equation}\label{eq:angularresolution}
    \sigma_\theta(E_r) = p_\theta \sqrt{\frac{E_{\rm ref}}{E_r}} \, .
\end{equation}
We interpret $p_\theta$ as the angular resolution at some reference energy $E_{\rm ref}$. We choose this reference energy by finding the 10th percentile of the recoil energy distribution from solar neutrinos above threshold. This corresponds to $\sim2$~\keVr for nuclear recoils and $\sim35$~keV for electron recoils (recall that we set hard thresholds of 1 keV and 10 keV on nuclear and electron recoils respectively). In this way, we can interpret $p_\theta = 5^\circ$ as implying that 90\% of all recoil directions could be measured with \emph{at worst} a 5-degree accuracy. From comparisons with prior results~\cite{Vahsen:2020pzb}, we find that this fit is relatively close to other more complicated functional forms.
We additionally prevent values from exceeding $\sigma_\theta>57.3^\circ$ which have no meaning. In Fig.~\ref{fig:PerformanceParameters_ERs}, we show some examples of the equivalently-defined angular resolution parameters obtained by attempting track reconstruction on simulated electron recoils in atmospheric He and CF$_4$ mixtures~\cite{SamueleThesis,CYGNO_inprep}. We take these as inspiration for our performance benchmarks.

\subsubsection{Energy resolution}
The fractional energy resolution will also improve with increasing energy. A functional form from Ref.~\cite{Vahsen:2020pzb} is \mbox{$\sigma_E/E = \sqrt{\left(a^{2} / E^{2}+b^{2} / E+c\right)}$}, but for the same reasons as stated above we will adopt the simplified form,
\begin{equation}\label{eq:energyresolution}
    \frac{\sigma_E(E_r)}{E_r} = p_E \sqrt{\frac{E_{\rm ref}}{E_r}} \, .
\end{equation}
We take the same reference energies as for the angular resolution, and interpret $p_E$ as the fractional energy resolution that could be obtained by at least 90\% of observed recoils.

To simplify our discussion, we have defined three performance benchmarks ranging from an optimistically good performance to a more realistic performance. The energy-dependence of the energy resolution, angular resolution and head-tail efficiency are displayed in Fig.~\ref{fig:PerformanceParameters_ERs}. Finally, in Fig.~\ref{fig:SmearedDistributions} we show how the distributions of electron and nuclear energies and track angles change once we have applied all of these performance effects. These three binned event distributions will constitute our inputs to the sensitivity study in the final section.

\subsection{Track lengths}\label{sec:tracklengths}
\begin{figure}
\begin{center}
\includegraphics[trim = 0mm 0 0mm 0mm, clip, height=0.3\textwidth]{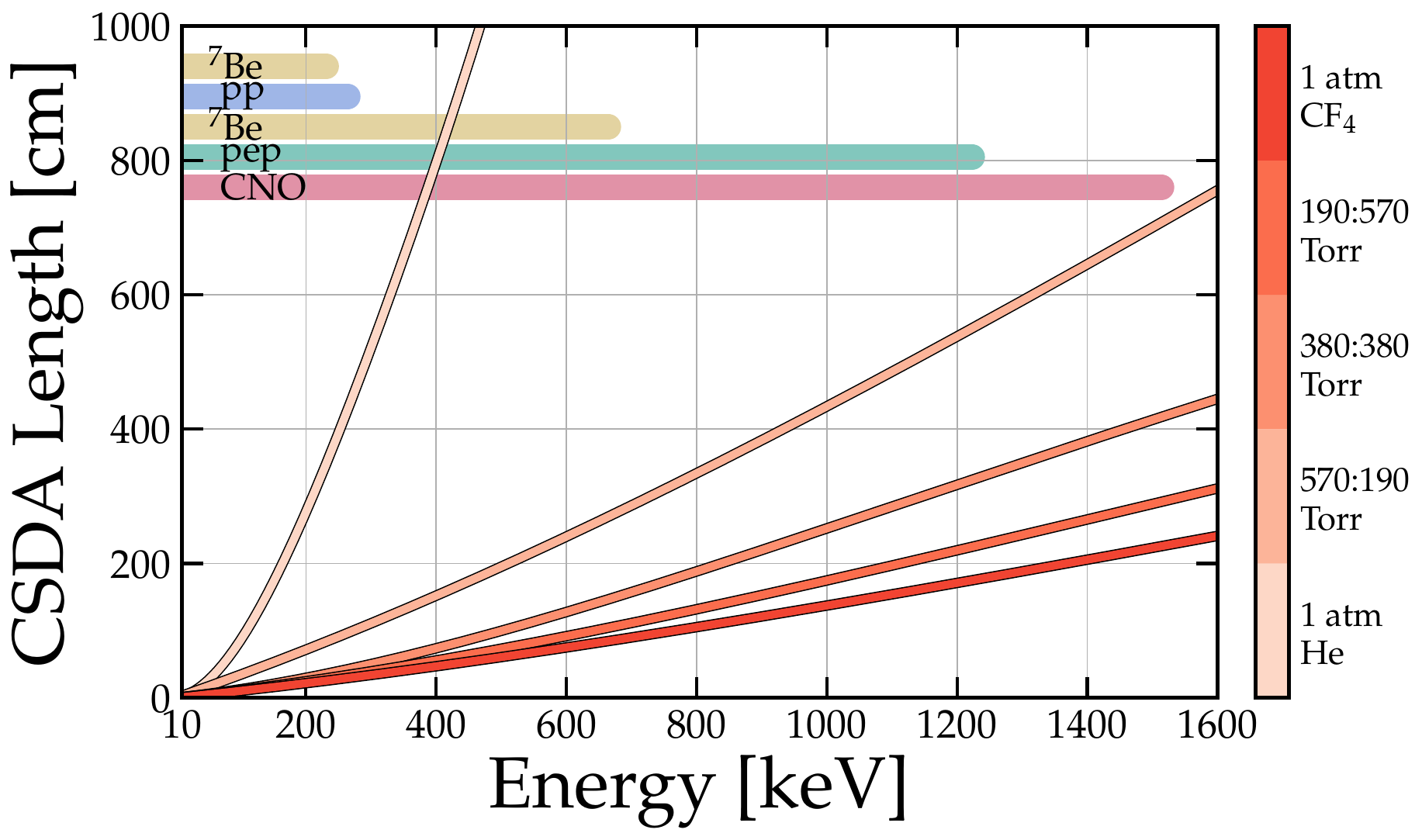}
\caption{Expected recoil track lengths as estimated using electron CSDA range calculated by NIST's \text{estar} program~\cite{148751}. The lengths displayed correspond to the different mixtures of He:CF$_4$ at 1 atmosphere shown along the orange line of Fig. \ref{fig:GasDensity_vs_Events}. The colour scale encodes the different mixtures and is darker for higher concentrations of CF$_4$. On the top left, we also display the maximum recoil energy caused by a sample of neutrino sources. These are colour-coded and utilise the same colour scheme as Fig. \ref{fig:NeutrinoFluxes}.} 
\label{fig:LengthVEn}
\end{center}
\end{figure}

\begin{figure}
\begin{center}
\includegraphics[trim = 0mm 0 0mm 0mm, clip, height=0.45\textwidth]{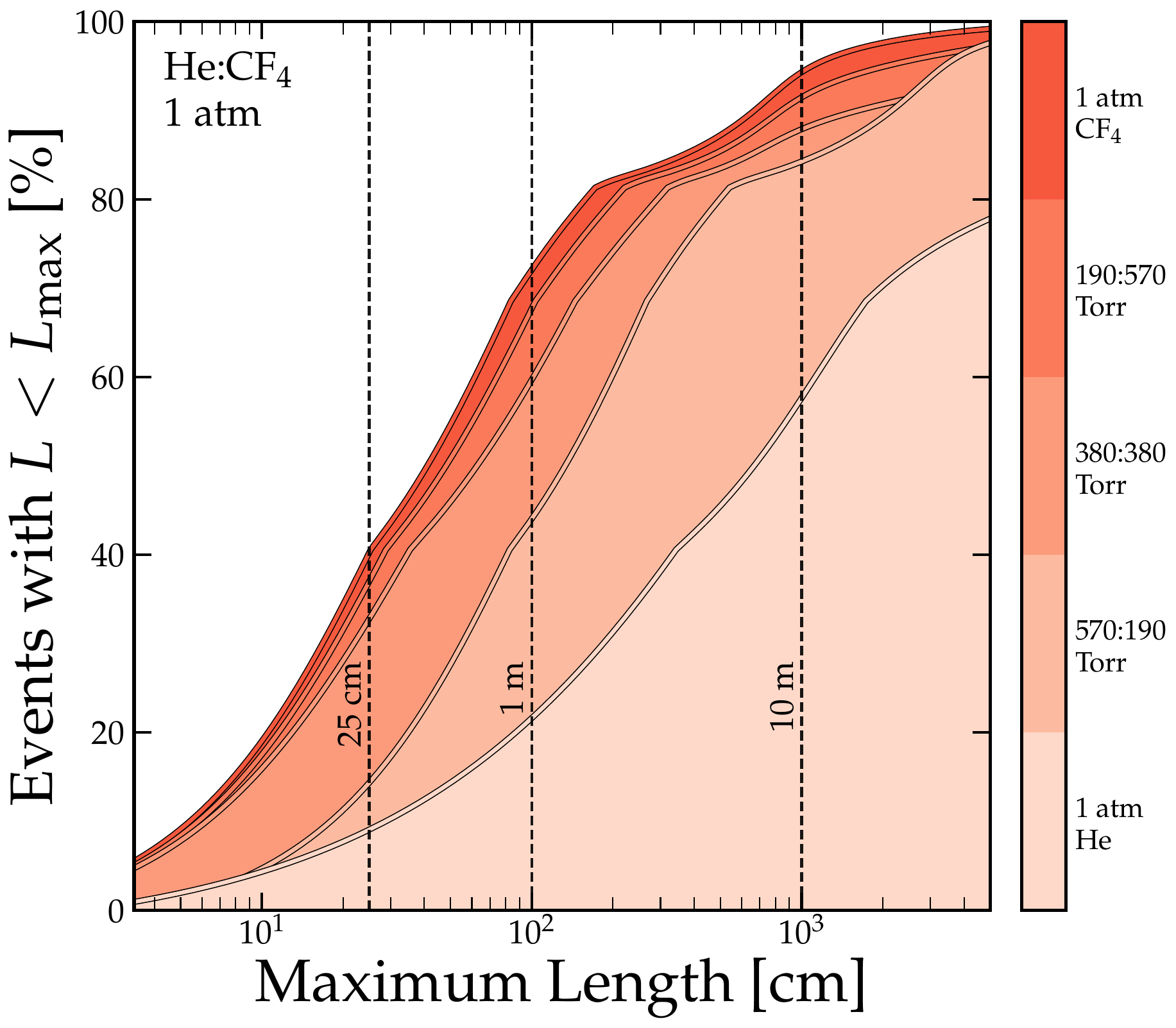}
\caption{Fraction of electron tracks in 1 atmosphere of He:CF$_4$ whose length $L$ is less than some maximum cutoff $L_\text{max}$. The colour scheme is the same as in Fig. \ref{fig:LengthVEn} and encodes the amount of CF$_4$ relative to He. As the concentration of CF$_4$ increases, more of the tracks are contained inside the TPC volume. We mark on the plot a few relevant dimensions, such as the approximate side-lengths of a 10 m$^3$ TPC module, as well as the mean drift length of 25~cm. It is important to emphasise here that this length reflects the straight-line CSDA range and not the physical size of the track, and so is an extreme upper limit on the maximum size of an electron track, which in reality will follow extremely non-linear trajectories}
\label{fig:fractionvminlength}
\end{center}
\end{figure}

One concern that arises when considering electron recoil tracks in gas out to energies of several hundred keV is their \textit{lengths}. We are assuming \Cygnus will consist of modules with $\sim$10 m$^3$ volume, but to obtain good directionality the dimension of the drift direction must be kept to around 50~cm at most. This is to limit the diffusion of the charge which will wash out the directionality, and keeping this length short is non-negotiable. Unfortunately, the range of electrons in atmospheric pressure gas can reach several metres for the highest energies we consider here. To obtain accurate measurements of the recoil energy and the direction we require the majority of the tracks to be contained in the volume, so we pause now to consider whether we can safely assume this will be the case for our proposal.

An upper limit on the expected electron track lengths can be estimated using the Continuous Slowing Down Approximation (CSDA). This range approximates the length travelled by a particle as it slows down to rest by taking the integral of the particle's stopping power with respect to energy~\cite{148751}. We emphasise that this approximation will give us an estimate of the straight-line length of the track, as opposed to the physical size of the track---the latter is more pertinent for this discussion but would require performing simulations of electrons over the full energy range and for a wide range of gas mixtures.\footnote{This is already being done in the context of \Cygnus and CYGNO (see e.g.~\cite{SamueleThesis,Schueler:2022lvr,Ghrear:2020pzk,Ghrear:2024rku}, but is not the focus of our study.} In reality, these tracks will be highly non-linear---the electrons can double back on themselves multiple times as they recoil through the medium. This would cut their total physical size down by a factor of two or more from the naive CSDA range. Nonetheless, since our discussion here is about how detrimental these excessively long tracks could be to our sensitivity, the fact that the CSDA gives us an overestimate means we remain conservative. We focus on our baseline gas mixture of He:CF$_4$ at one atmosphere with a variable ratio of helium.

We use NIST's \textit{estar} program to calculate the CSDA range~\cite{148751} for an electron in the different mixtures, with the results shown in Fig.~\ref{fig:LengthVEn}. As would be expected, a higher concentration of CF$_4$---i.e.~a higher density gas---is desirable if our goal is to successfully contain all of the electron tracks within a single module with a maximum 10-metre-scale linear dimension.

Using this CSDA range, we then show the fraction of recoil events with lengths \textit{shorter} than some maximum cutoff in Fig.~\ref{fig:fractionvminlength}. At the very worst, for 1 atmosphere of He, around 60\% of the tracks would be longer than 10 m. If we assume we have a $\sim$10~m~$\times$~10~m~$\times$~1~m TPC module in a back-to-back design, as in Ref.~\cite{Vahsen:2020pzb}, this could potentially lead to a sizeable fraction of tracks being only partially contained. However, these lengths shrink with higher fractions of CF$_4$, and for a roughly 50:50 mixture, we have only 10\% of the very highest-energy tracks with straight-line lengths that are longer than the side length of the TPC module.

We reiterate that this is still a worst-case scenario, the majority of tracks will be substantially smaller in size than the naive CSDA range. Although it is still plausible that some fraction of the highest energy events may extend partially outside of the fiducial gas volume, these results suggest that this will not be a huge number. In any case, with advanced machine-learning-powered algorithms for track reconstruction (see e.g.~Ref.~\cite{Ghrear:2024rku}) we may even be able to tolerate such cases and still obtain reasonably accurate energy and angle estimates from the track segments that exist inside the fiducial volume. That said, it is still worth pointing out that the 200--1400 keV energy range is the primary region of interest for CNO neutrinos, meaning this issue will be worth a dedicated simulation study in the context of a 1000-m$^3$-scale \Cygnus experiment.

\section{Neutrino reconstruction}\label{sec:nureconstruction}
\begin{figure}
\begin{center}
\includegraphics[trim = 0mm 0 0mm 0mm, clip, width=0.49\textwidth]{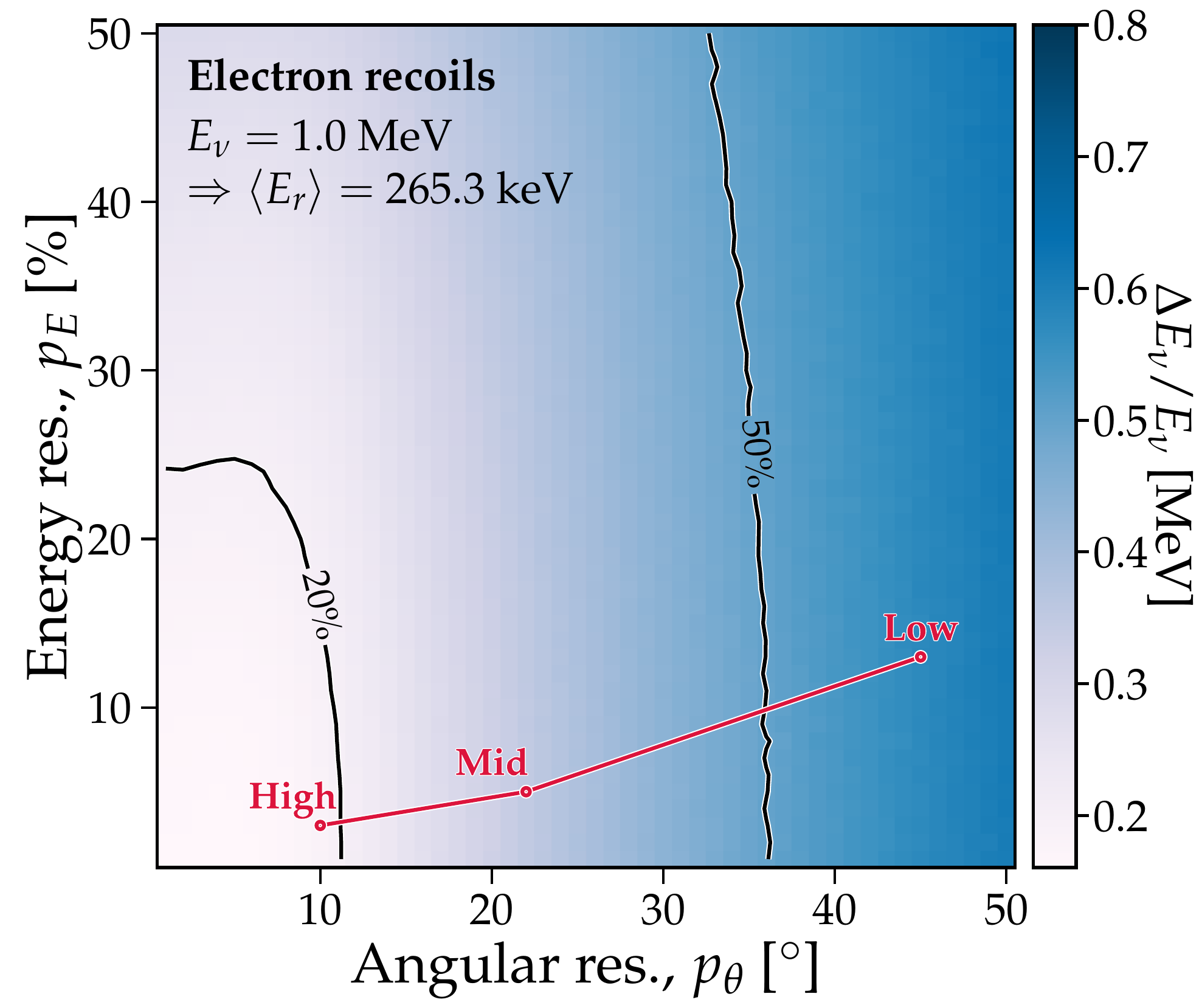}
\includegraphics[trim = 0mm 0 0mm 0mm, clip, width=0.49\textwidth]{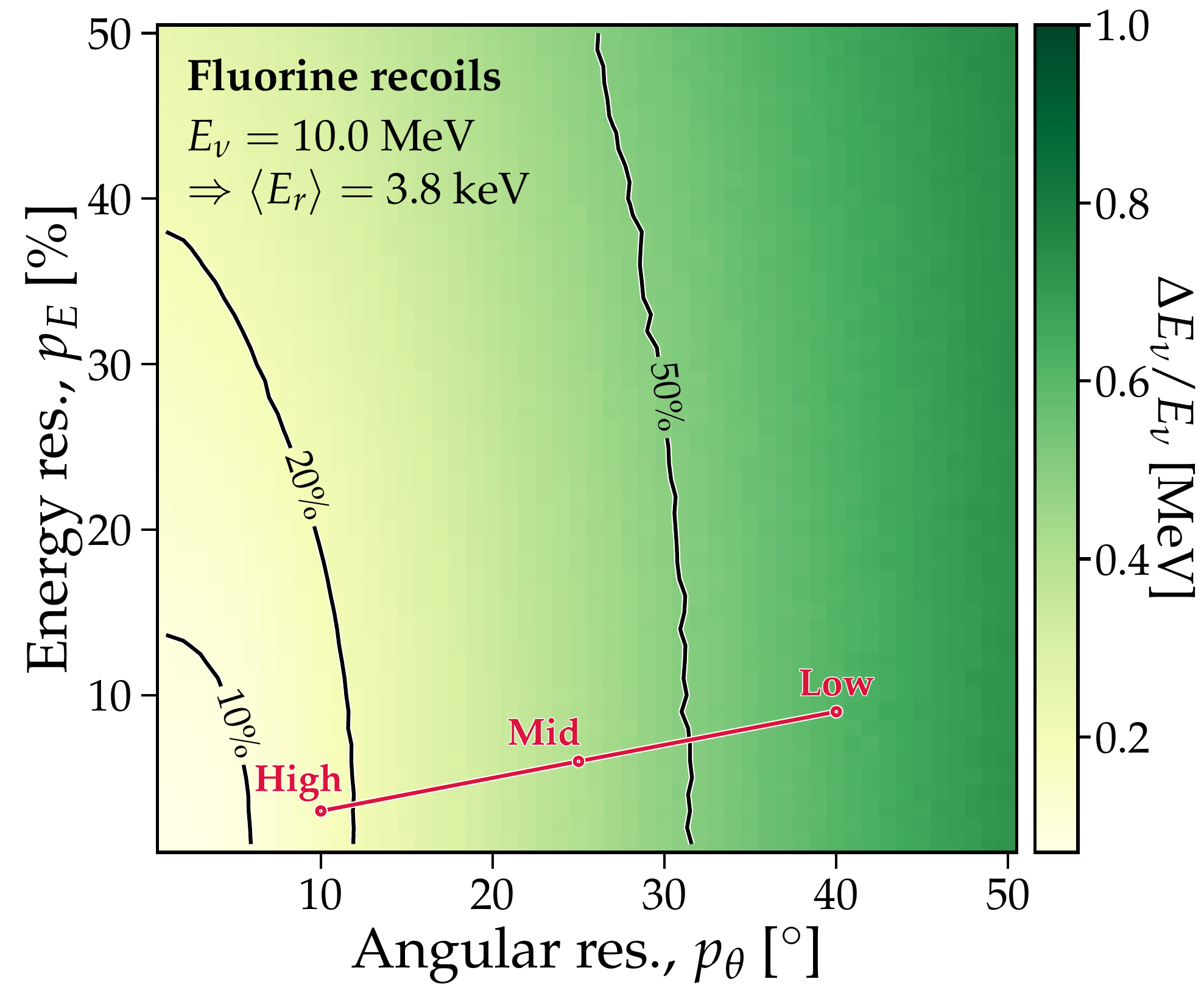}
\caption{Neutrino energy reconstruction accuracy as a function of energy and angular resolution. The colour scale encodes how accurately an initial neutrino energy of 1 MeV (top panel) or 10 MeV (bottom panel) could be reconstructed from electron (top) and nuclear (bottom) recoil energy and directions. The horizontal axis describes the angular resolution parameterised according to Eq.(\ref{eq:angularresolution}), and the vertical axis describes the fractional energy resolution parameterised according to Eq.(\ref{eq:energyresolution}). We define $\Delta E_\nu$ to be the size of the 68\% region around the median reconstructed neutrino energy. We also highlight three ``benchmark'' performance scenarios (high, medium and low) that were introduced in Fig.~\ref{fig:PerformanceParameters_ERs} and which we will use to simplify the discussion in later examples.} 
\label{fig:NuEnergyResolution}
\end{center}
\end{figure}

Before we quantify \Cygnus's sensitivity in terms of flux measurements, we can gain some further insight into how our benchmark energy and angular resolution choices limit the ability of the detector to reconstruct the initial neutrino energy. Using the parameters $p_\theta$ and $p_E$ that we defined in the previous section, we can vary these and determine the accuracy $\Delta E_\nu$, with which a monoenergetic neutrino source's energy could be reconstructed. We do this by creating a large sample of Monte Carlo generated recoils for a given set of $p_\theta$ and $p_E$, and then use Eq.(\ref{eq:kinematics}) to determine the neutrino energy. We then define $\Delta E_\nu$ as half the width of that distribution between its 
18th and 84th percentiles. We use this rather than the full width at half maximum, because the resulting distribution of inferred $E_\nu$ is typically highly asymmetric, with a large tail for $E_\nu>E^{\rm true}_\nu$

The result of this exercise is shown in Fig.~\ref{fig:NuEnergyResolution} for electron and nuclear recoils. In the former case, we use an initial neutrino energy of \mbox{$E_\nu = 1$~MeV}, i.e.~a typical CNO neutrino energy, whereas for the nuclear recoil case, we take a 10 MeV energy typical of \Boron neutrinos. As one would expect, the fractional reconstruction accuracy $\Delta E_\nu/E_\nu$ becomes worse as both $p_E$ and $p_\theta$ are decreased. Obtaining better than 10\% accuracy for the majority of events requires excellent performance, i.e. better than 10\% energy resolution and 5$^\circ$ angular resolution. 

These performance goals seem daunting, however it should be mentioned that this is only the accuracy with which a single neutrino's energy could be reconstructed. In practice, we will have a distribution of events, and as in Fig.~\ref{fig:NuFlux_reconstruction}, we should be able to average out these individual errors in a distribution that can be compared with expected flux models. Moreover, recall that experiments without any directional sensitivity cannot do this event-by-event energy reconstruction \emph{at all}. Notice that the error shoots up to over 100\% at the edges of Fig.~\ref{fig:NuEnergyResolution}, so even a 50\% error is a significant improvement over the non-directional case.

To proceed to the next section where we will examine how this performance translates into measurements of the actual fluxes, we need to make our discussion slightly more concise. To do this we will collapse the result from Fig.~\ref{fig:NuEnergyResolution} down into three performance benchmarks. We label these in terms of their level of neutrino reconstruction accuracy: ``high'', ``mid'' and ``low''. It should be noted that these are aimed simply to cover a suitable range of capabilities, with the intention that results from experiments or gas simulations could be mapped onto these scales. Nevertheless, we anticipate that our High-performance benchmark will be a significant challenge, whereas the Mid-range performance is simply optimistic, and the Low performance is realistic and already surpassed using simple track reconstruction algorithms in simulations for CYGNO~\cite{CYGNO:2023ucc,SamueleThesis} and \Cygnus~\cite{Schueler:2022lvr,Ghrear:2020pzk,Ghrear:2024rku}.

\subsection{Fitting neutrino fluxes}
\begin{figure*}
\begin{center}
\includegraphics[trim = 0mm 0 0mm 0mm, clip, height=0.45\textwidth]{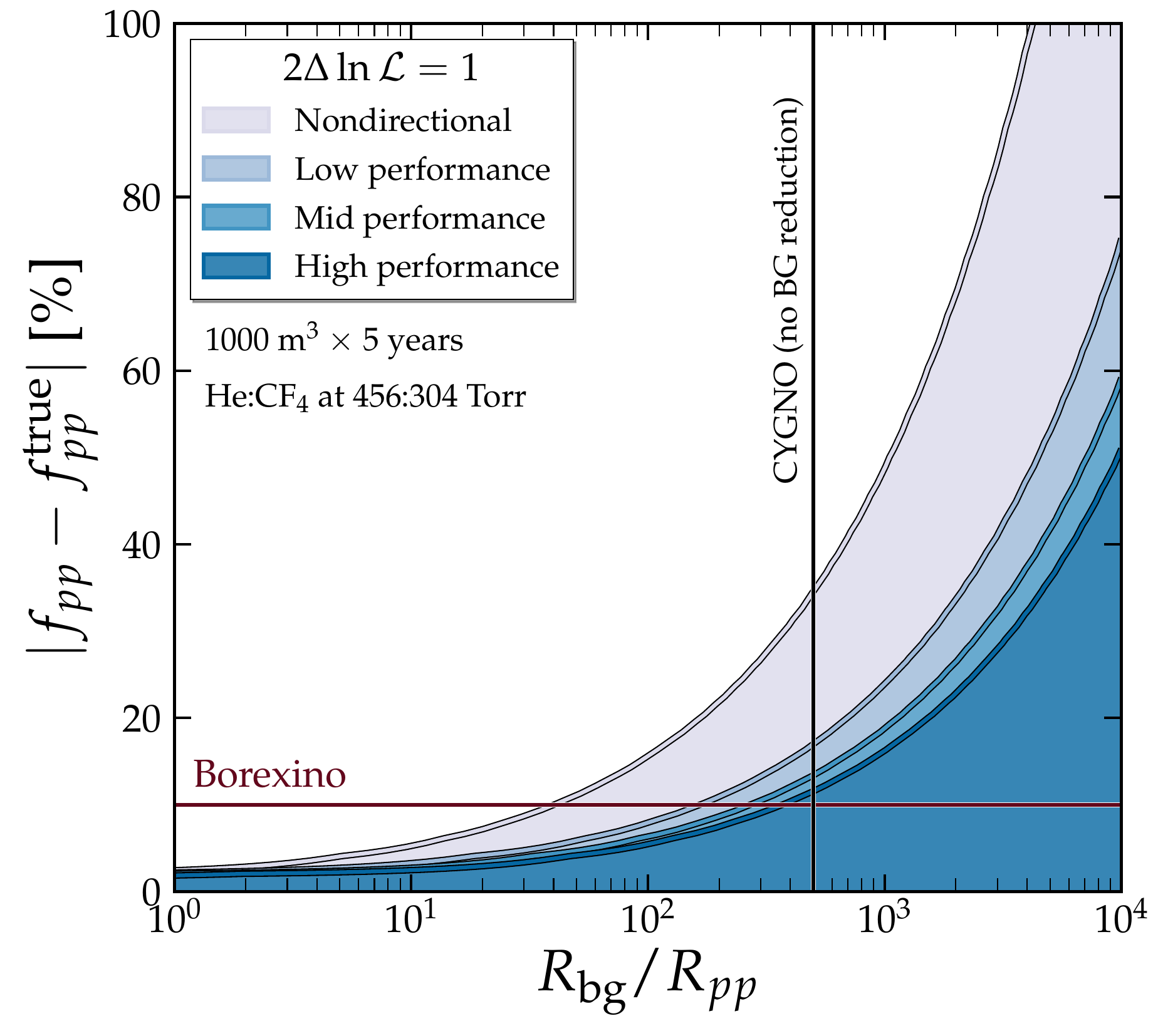}
\includegraphics[trim = 0mm 0 0mm 0mm, clip, height=0.45\textwidth]{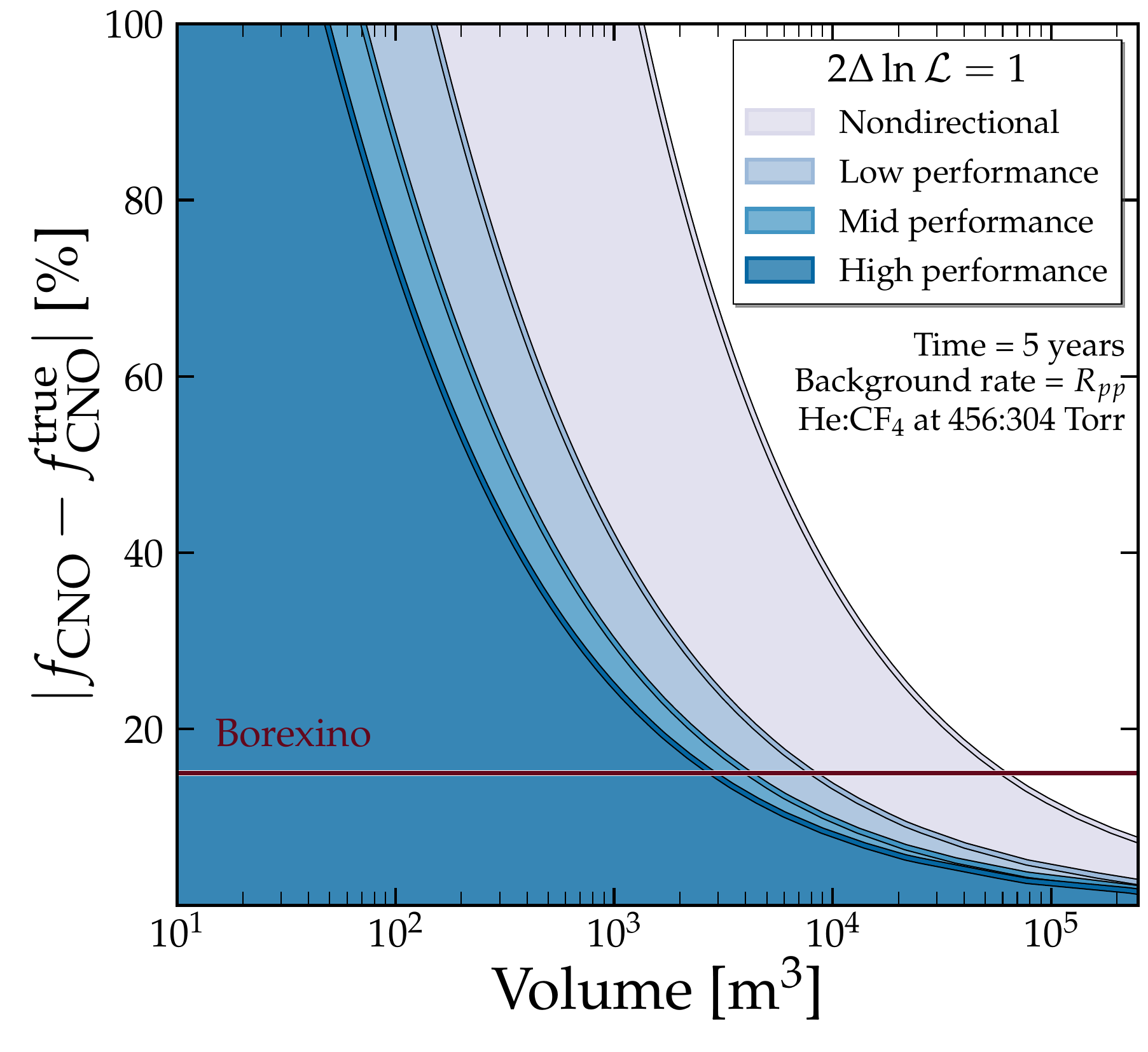}
\caption{Error on the reconstructed flux normalisations of $pp$ (left) and CNO neutrinos (right) normalised to the B16 high-Z standard solar model of Ref.~\cite{Vinyoles:2016djt}. We display the estimated median $1\sigma$ measurement as a function of TPC volume for a fixed non-neutrino background rate on the right, and as a function of the non-neutrino background rate for a fixed volume on the left. We do this because we anticipate that positively detecting the $pp$ flux will be more demanding of the low background conditions, whereas detecting CNO neutrinos will be more demanding of the directionality, as we can see in the shape of the various contours that correspond to different scenarios ranging from the best directional performance (darkest colour), to the worst directional performance (lightest colour). We emphasise that our limit labelled `Nondirectional' has exactly the same energy resolution and efficiency as the \emph{best} directional limit shown, thereby proving the value of incorporating directional information.} 
\label{fig:NuFluxFit_1D}
\end{center}
\end{figure*}

\begin{figure*}
\begin{center}
\includegraphics[trim = 0mm 0 0mm 0mm, clip, height=0.44\textwidth]{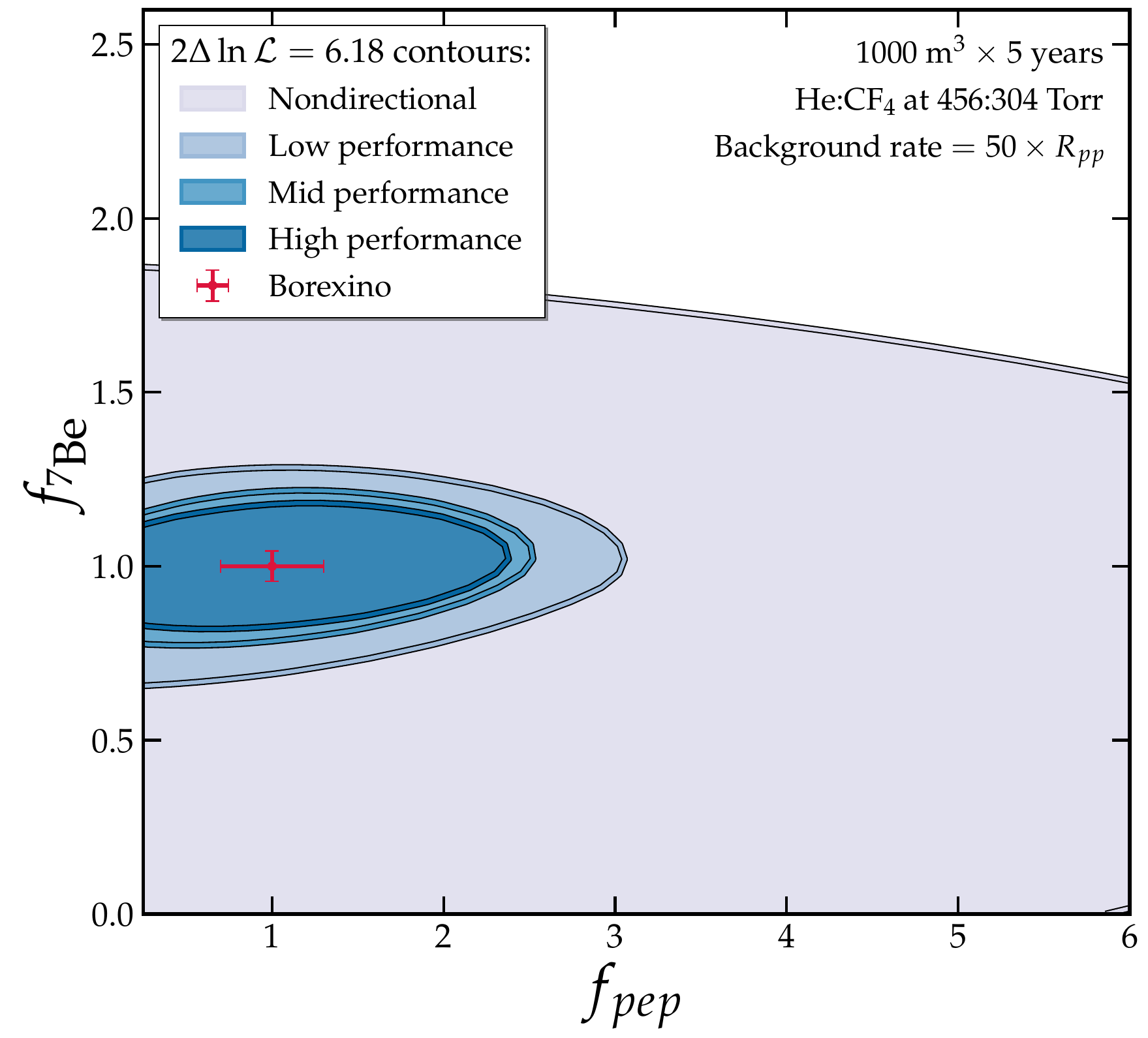}
\includegraphics[trim = 0mm 0 0mm 0mm, clip, height=0.44\textwidth]{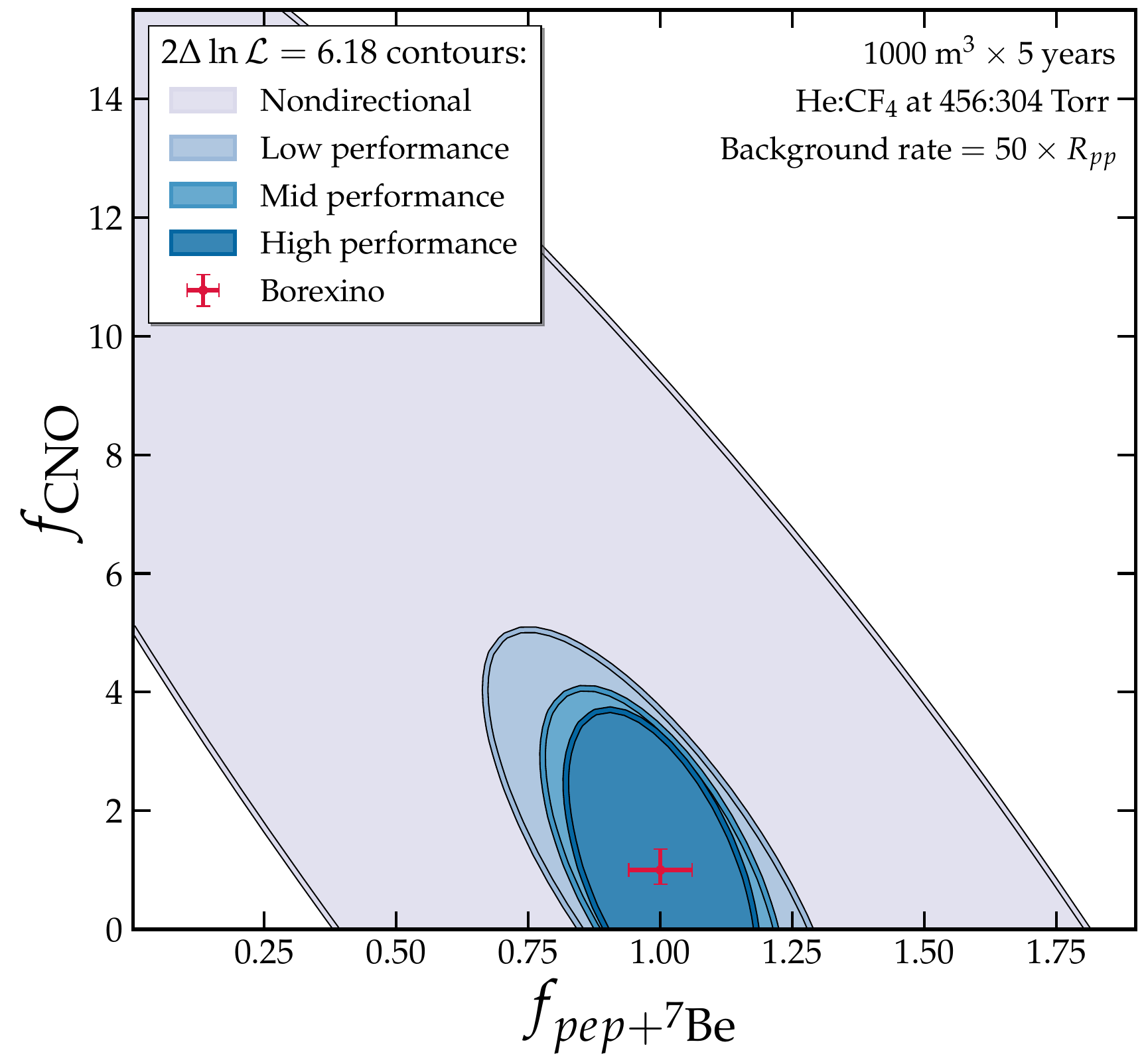}
\caption{Projected median 2$\sigma$ sensitivity to neutrino flux parameters normalised to the B16 high-Z standard solar model of Ref.~\cite{Vinyoles:2016djt}. The various colours correspond to different levels of directional sensitivity, ranging from a best-case scenario based on optimistic projections for electron recoil energy/angular resolution from gas simulations (darkest blue), down to a worst-case scenario with no directional sensitivity at all (lightest blue). The left-hand panel shows the expected sensitivity to the $^7$Be and $pep$ fluxes that overlap in the $\sim$200 keV recoil energy region, whereas the right-hand panel shows the expected 2$\sigma$ sensitivity to a combined measurement of the CNO and $pep$+$^7$Be fluxes. In both cases we fix the background rate to be equal to the $pp$ electron recoil rate. As in the previous figure, we again see that the directionality greatly improves the capability of the experiment to measure the fluxes, simply due to the added background rejection, and the additional handle to separate the different neutrino fluxes from each other. In particular, we point out the \textit{shapes} of the contours, where the negative correlation between the CNO and neutrino lines can be partially reduced in the best-performing case (darkest blue).} 
\label{fig:NuFluxFit_2D}
\end{center}
\end{figure*}
To project sensitivity to the various parameters controlling the neutrino fluxes we will perform a profile likelihood ratio test~\cite{Cowan:2010js}. This allows us to much more straightforwardly account for all of the detector effects discussed so far, as well as the presence of a non-neutrino background.

First assuming that we have a model for the spectra of the neutrino signal components as well as a background, we can construct a parameter space $\boldsymbol{f} = \{f_{\nu_1},...f_{\nu_n},f_{\rm bg} \}$ which control the scale of each one in terms of a series of constants. We can treat each of the $n$ neutrino fluxes separately, or we can combine them together for instance $f_{\rm CNO} = f_{^{13}{\rm N}}+f_{^{15}{\rm O}}+f_{^{17}{\rm F}}$. In all of the examples we present in this section, the sensitivity is driven by the electron recoils as opposed to nuclear recoils, which have a subdominant rate and do not influence the flux measurements to which we will project our sensitivity. Hence, we do not impose any assumption about nuclear or electron recoil discrimination, and instead only focus on neutrino events versus non-neutrino events. 

For reasons that we will explain below, we adopt a binned likelihood for our pseudo-data. The bins run over the space $(E_r,\cos{\theta_\odot})$. Given an observed number of events in each bin $N^{ij}_{\rm obs}$ and an expected number $N^{ij}_{\rm exp}(\boldsymbol{f})$, the likelihood is constructed as the product of the Poisson probability distribution function ($\mathscr{P}$) in each bin. An option we have is to multiply the likelihood by a series of Gaussian distributions for each parameter that reflect our prior knowledge about each parameter. However, since we are aiming to determine the accuracy with which we can measure each parameter in $\boldsymbol{f}$, we choose not include these. Since our background rate is the only nuisance parameter that we are not interested in, we will only include one of these additional functions. The full likelihood is then simply,
\begin{equation}\label{eq:likelihood}
 \mathscr{L}(\boldsymbol{f})= \prod_{i,j} \mathscr{P} \left(N_\textrm{obs}^{ij} \bigg| N^{ij}_{\rm exp}(\boldsymbol{f})\right) \, ,
\end{equation}
where the expected number of events in bin $i,j$ is,
\begin{align}
    N_{{\rm exp}}^{i j}(\boldsymbol{f})=M T & \int_{\cos \theta_{\odot}^{i}}^{\cos \theta_{\odot}^{i+1}} \int_{E_{r}^{j}}^{E_{r}^{j+1}}\left[\frac{\mathrm{d}^{2} R(\boldsymbol{f})}{\mathrm{d} E_{r} \mathrm{~d} \cos \theta_{\odot}}\right] \nonumber \\
    & \mathrm{~d} \cos \theta_{\odot} \mathrm{~d} E_{r} \, ,
\end{align}
where,
\begin{align}
    \frac{\mathrm{d}^{2} R(\boldsymbol{f})}{\mathrm{d} E_{r} \mathrm{~d} \cos \theta_{\odot}} &= f_{\rm bg} R_{\rm bg} \frac{\varepsilon(E_r)}{\Delta\cos{\theta_\odot}\Delta E_r \int_{E_{\rm min}}^{E_{\rm max}} \varepsilon(E_r)\textrm{d}E_r} \nonumber \\
    &+ \sum_{i=1}^n 2\pi f_{\nu_i} \frac{\mathrm{d}^{2} R_{\nu_i}}{\mathrm{d} E_{r} \mathrm{d}\Omega_q} \, ,
\end{align}
and $R_{\rm bg}$ is the background rate in units of (ton-year)$^{-1}$. In other words, we are taking the most conservative assumption that the background is completely isotropic and flat in energy and that its background normalisation is floated in the fit in the most uninformed way, i.e.~with no additional component in the likelihood function that expresses our prior knowledge of it, as is usually be implemented in real analyses. We will assume that $R_{\rm bg}$ is larger than the $pp$ neutrino rate. Preliminary simulations of internal backgrounds (from e.g.~the GEMs, field cage, cathode, cameras, lenses etc.) for a 30-m$^3$ CYGNO optical TPC are at the level of $\sim 10^4$ events per year~\cite{SamueleThesis,CYGNO_inprep}, which is a factor of $\sim$500 larger than $R_{\rm pp}$. It is important to emphasise that these backgrounds are based on the existing measurements of the radiopurity of components inside MPGD elements, which are typically much higher than would usually be tolerated in a low-background experiment. Dedicated screening efforts within DRD1 at CERN over the next few years are anticipated to lead to significant reductions from this worst-case scenario~\cite{Surrow:2022ptn}. We will therefore show some results where we assume a background reduction of a factor of 10 is achieved, i.e. $R_{\rm bg} = 50 \times R_{\rm pp}$.

Using the likelihood written above, we can construct a hypothesis test statistic out of a maximum likelihood ratio, which compares the maximised likelihoods between two values. The interesting case for us will be to test for the presence of the neutrino signal and exclude the null hypothesis that the signal is absent, i.e. one for $f_0 \in \boldsymbol{f}$ we can test against $f_0 = 0$. The likelihood ratio of interest here will then be,
\begin{equation}\label{eq:likelihood-ratio}
\Lambda = \frac{ \mathscr{L} (0,\hat{\hat{\boldsymbol{f}}}) }{\mathscr{L} (\hat{f}_0,\hat{\boldsymbol{f})}}\, ,
\end{equation}
where~$\mathscr{L}$ is maximised at $\hat{\hat{\boldsymbol{f}}}$ when $f_0 = 0$ and $\hat{\boldsymbol{f}}$ when $f_0$ is a free parameter.

We define the test statistic for this likelihood ratio as,
\begin{equation}
	{\rm TS} = \left\{ \begin{array}{rl}
	-2\ln \Lambda  & \, \,  f_0>0  \,,\\
	0  & \, \, f_0\le 0, \, \,  \,.
	\end{array} \right. 
\end{equation}
Since the two models differ by the fixing of one parameter, and our null hypothesis has a parameter set to the boundary of the allowed space, Chernoff's theorem~\cite{Chernoff:1954eli} holds. This is a generalisation of Wilk's theorem in which the statistic $q_0$ is asymptotically distributed according to \mbox{$\frac{1}{2}\chi^2_{1}+\frac{1}{2}\delta(0)$} when the $\mathcal{M}_{\sigma=0}$ hypothesis is true. The practical consequence of this for us is that the significance of the signal tested against the $f_0 = 0$ hypothesis is simply $\sqrt{\rm TS}$. See Ref.~\cite{Cowan:2010js} for a detailed discussion of the use of these asymptotic formulae.

The distribution of TS would normally be calculated using many Monte Carlo realisations of pseudo-data. However, there is a trick we can use to greatly reduce this computational expense. We can instead instantly calculate the expected (median) discovery limit using the Asimov dataset~\cite{Cowan:2010js}. This is a hypothetical scenario in which the observation exactly matches the expectation for a given model, i.e.~$N^{ij}_{\rm obs} = N^{ij}_{\rm exp}$ for all $i$ and $j$. It can be shown that the test statistic computed assuming this dataset asymptotes towards the median of the model's TS distribution as the number of observations increases~\cite{Cowan:2010js}.

We use the statistical methodology described above to evaluate the prospects for measuring various neutrino fluxes. We will display our results as a function of the three detector performance benchmarks, as discussed in the previous section. We fix our gas mixture to our baseline of 60:40 ratio of He:CF$_4$ at one atmosphere, because our ``Mid'' performance benchmark has already been shown to be feasible with this gas density in simulation studies for CYGNO~\cite{CYGNO_inprep, SamueleThesis}. It would be simple to achieve improved sensitivity estimates just by assuming higher fractions of CF$_4$ than this, however, we must remember that higher densities will always incur a loss in directional performance.

We now come to our results. First, we consider the measurement of individual fluxes, using the existing measurements by Borexino as a point of comparison. Again, we do not anticipate that \Cygnus will be more competitive than Borexino, as it is smaller-scale and intended to be a multi-purpose instrument, but a comparison against Borexino allows us to appreciate how we match up against the state-of-the-art. 

Our projected median 1$\sigma$ constraints on the flux of $pp$ neutrinos, and the flux of CNO neutrinos are shown in Fig.~\ref{fig:NuFluxFit_1D}. For the $pp$ flux we expect this to be measurable with smaller total volumes, however the accuracy of this measurement will depend strongly on the size of the non-neutrino background that it is effectively competing with. Therefore, for the left-hand panel of Fig.~\ref{fig:NuFluxFit_1D} we show the error on the reconstructed flux as a function of the background rate relative to the $pp$ rate, $R_{\rm bg}/R_{pp}$. As expected, the better the directional performance, the better the estimate of the flux. In fact, we find that if the detector has very good angular and energy resolutions, the measurement in a 1000 m$^3$ total volume would be competitive with Borexino's 10\% measurement of the $pp$ flux, even if there were a factor of several hundred times more background events than neutrinos. Given internal background simulations for CYGNO~\cite{CYGNO_inprep} which suggest $R_{\rm bg}/R_{\rm pp}\sim 500$ is already achievable without any dedicated background reduction, we deem this to be quite promising. The difference between the lightest blue contour in Fig.~\ref{fig:NuFluxFit_1D} and the dark blue ones emphasises the importance of the directionality, even in this simple example which is mostly a case of identifying signal events over an isotropic background.

Next, in the right-hand panel of Fig.~\ref{fig:NuFluxFit_1D}, we consider the error in the measurement of the CNO flux. The key factor in this case will be simply acquiring enough signal events, so instead we display the median 1$\sigma$ error on the reconstructed flux as a function of volume. We see an even more dramatic dependence on the detector performance here. This is to be expected, since measuring the CNO flux requires distinguishing it from other fluxes, which is extremely challenging using just recoil energy information alone. We find that a 1000 m$^3$ experiment can already constrain the CNO flux to within 20--30\% whereas a non-directional experiment cannot even distinguish it from zero at 1 $\sigma$. As stated earlier, this sensitivity could stand to improve with more CF$_4$, but there would be some trade-off with the poorer directional sensitivity in higher-density gases. However, we point out that in the case of CNO neutrinos the loss of directionality may not be as severe since the relevant energy range is around 200--1400 keV. In this window, the tracks are significantly longer, and so their reconstruction could potentially tolerate a higher level of diffusion. So while this measurement would be challenging, it does not seem completely out of the question for \Cygnus-1000.

As we hinted at above, the measurement of the CNO flux is limited by the detector's ability to distinguish the various fluxes from each other. So finally, in Fig.~\ref{fig:NuFluxFit_2D} we show the joint measurement of two flux contributions together, again as a function of detector performance. We specifically choose measurements that are expected to be challenging without directional sensitivity, due to the overlapping nature of the different fluxes. The left-hand panel shows the measurement of the $^7$Be and $pep$ neutrino lines, and the right-hand panel shows the combined measurement of the CNO flux and $pep$+$^7$Be lines. The most notable takeaway from this figure is the near-disappearance of the negative correlation in the right-hand panel for the best performance benchmark. Without directional sensitivity, there is a strong degeneracy between the CNO flux and the neutrino lines it is hidden underneath. For the best directional performance, not only can the fluxes be measured more precisely but the degeneracy between different fluxes be alleviated substantially.

\section{Conclusions}\label{sec:conc}
In this article, we have quantified the extent to which gaseous time projection chambers under the umbrella of the \Cygnus Consortium could perform as detectors of solar neutrinos. Although the primary goal of \Cygnus is to push the detection of dark matter into the neutrino fog, we have shown that there is scope to make interesting measurements of the directions of solar-neutrino-induced electron recoils, even in smaller scale stepping-stone experiments. These results are timely given that plans for the next stage of high-definition-readout \Cygnus prototypes at the 10--30~m$^3$ scale are under development right now~\cite{OHare:2022jnx, Surrow:2022ptn}.

Our primary goal was to evaluate the background conditions and directional performance necessary to achieve certain levels of precision when it comes to the measurement of the solar neutrino flux. Firstly, regarding the background, the relevant comparison is with the rate of $pp$ neutrinos, which for our benchmark target of 456:304 Torr of He:CF$_4$ yields a rate of around 10 events per year per 10 m$^3$ of volume (or larger if higher gas densities can be tolerated). Thanks to directional sensitivity, a measurement of the $pp$ neutrino flux using these events is possible even when background conditions generate an event rate across the same energy window that is several hundred times larger than the signal. More concretely a 1000~m$^3$ experiment with good directionality (i.e. $\sim$20--10$^{\circ}$ angular resolution below 100 keV) can measure the $pp$ flux to the 10\% level with a background that is 200-300 times that of the rate of $pp$ events. A 10~m$^3$ experiment run for 3 years would be sufficient for an initial discovery of the signal assuming a factor of 10 background reduction from preliminary simulations of the internal backgrounds in a directional TPC using an optical readout~\cite{SamueleThesis,CYGNO_inprep}.

Regarding the other neutrino fluxes, we have seen, for example in Fig.~\ref{fig:NuFluxFit_2D}, that the directional information helps break degeneracies between different fluxes that overlap with one another in their recoil energy spectra. The demands placed on the gas density and the size of the experiment are higher in this case because the fluxes are lower, however, we have seen that even for 1000-m$^3$ scale experiments a combined constraint can be placed on neutrino fluxes like the CNO and $pep$ at the same time---a task that is essentially impossible in a non-directional experiment at the same scale.

The results from this study will be used when optimising \Cygnus and its prototypes over the next few years~\cite{Vahsen:2020pzb,OHare:2022jnx}, as the collaboration moves towards an ultimate design for a multi-purpose facility. It should also be noted that these results can be interpreted as benchmark performance requirements for other types of directional experiments making use of neutrino-electron recoils and can also be thought of as requirements to study neutrino-electron recoils originating from other sources of neutrino. Given that a shorter-term prototype TPC could be tested at a neutrino beam facility, our results suggest that such a detector could perform directional measurements of the neutrino scattering---a question we will devote to a future study.

\acknowledgments
We thank the members of the CYGNO collaboration for discussions and their support of this project. We also thank Nicole Bell and Jayden Newstead for comments on a draft of this paper. CL and CAJO are supported by the Australian Research Council under the grant number DE220100225. MG acknowledges support from the U.S. Department of Energy (DOE) via Award Number DE-SC0010504. VUB, LJB, FD, GJL, PCM and LJM
are supported by the Australian Research Council under grant number CE200100008.

\maketitle
\flushbottom

\bibliographystyle{bibi}
\bibliography{biblio}

\end{document}